\newcommand{\tsecompldate}{6th May 2004}
\documentclass[12pt,openbib]{article}
\usepackage{graphics}

\usepackage{amssymb}

\newcommand{\vol}[1]{\textbf{#1}}

\newcommand{\tpaptitle}[1]{``#1'',}
\newcommand{\tpretitle}[1]{``#1'',}
\newcommand{\tarttitle}[1]{``#1'',}
\newcommand{\tproctitle}[1]{in ``#1''}
\newcommand{\tinproctitle}[1]{``#1''}
\newcommand{\tbktitle}[1]{``#1''}

\newcommand{\tISBN}[1]{#1}

\newcommand{\tref}[1]{(\ref{#1})}
\newcommand{\fref}[1]{\ref{#1}}

\newcommand{\tnotpre}[1]{#1}

\newcommand{\tpre}[1]{}
\newcommand{\tprenote}[1]{}

\newcommand{\tnote}[1]{}
\newcommand{\href}[2]{#2}
\newcommand{\eprint}[1]{\texttt{#1}}
\newcommand{\tseurl}[1]{\texttt{#1}}

\newcommand{\tsedevelop}[1]{{}}

\renewcommand{\tnotpre}[1]{}
\renewcommand{\tpre}[1]{#1}
\renewcommand{\tprenote}[1]{\footnote{#1}}
\newcommand{\bea}{\begin{eqnarray}}
\newcommand{\eea}{\end{eqnarray}}
\newcommand{\beq}{\begin{equation}}
\newcommand{\eeq}{\end{equation}}

\newcommand{\Csymbol}[2]{\left(^{#1}_{#2}\right)}

\begin{document}

\renewcommand{\thefootnote}{\fnsymbol{footnote}}

%\tpre{
 \begin{flushright}
 Article for Contemporary Physics\\
 \texttt{Imperial/TP/3-04/12} \\
 \eprint{cond-mat/0405123} \\
 \tsecompldate \\
 \end{flushright}
 \vspace*{1cm}
%}

\begin{center}
{\Large\textbf{Complex Networks}}\tnote{tnotes such as this not
present in final version} \\
\tpre{\vspace*{1cm} }
 {\large T.S. Evans\footnote{email:
{\texttt{T.Evans AT imperial.ac.uk}}\tnotpre{,
\tsecompldate}\tpre{, WWW:
\href{http://theory.imperial.ac.uk/links/time}\texttt{http://theory.imperial.ac.uk/\symbol{126}time}
}}}
 \\
 \tpre{\vspace*{1cm}}
 \href{http://theory.imperial.ac.uk/}{Theoretical Physics},
 Blackett Laboratory, Imperial College London,\\
 Prince Consort Road, London, SW7 2BW,  U.K.
 \tnotpre{\\
 Tel: +44-(0)20-7594-7837,
 Fax: +44-(0)20-7594-7844 \\
 PACS:
 \\
 Key Words: }
\end{center}

%02.30.Cj Measure and integration
%02.30.Tb Operator theory
%11.10.-z  Field theory (for gauge field theories, see 11.15)
%11.10.Gh Renormalization

\begin{abstract}
An outline of recent work on complex networks is given from the
point of view of a physicist. Motivation, achievements and goals
are discussed with some of the typical applications from a wide
range of academic fields.  An introduction to the relevant
literature and useful resources is also given.
\end{abstract}

\renewcommand{\thefootnote}{\arabic{footnote}}
\setcounter{footnote}{0}

% ************************************************************
\subsection*{Introduction}

\begin{figure}[htb]
\begin{center}
  \scalebox{0.7}{\includegraphics{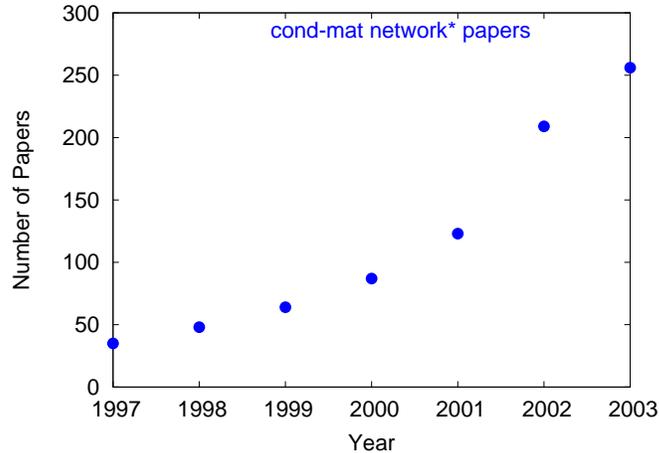}}
%  \scalebox{0.7}{\includegraphics{condmatnetworkhistoxls.eps}}
%  \scalebox{1.0}{\includegraphics{cmn2.eps}}
\end{center}
\caption{The number of papers listed per year on the
\texttt{cond-mat} archives with a word starting with ``network''
in their title.}
 \label{fcondmatpapers}
\end{figure}

These days, to gauge what is hot and what is not in the world of
physics, one need only turn to the electronic preprint archives.
If you look at the one focussed on condensed matter physics,
\texttt{cond-mat}, you will notice that among the preprints
looking at high temperature superconductivity, Bose-Einstein
condensation and other traditional pursuits of this community,
there are a large number talking about networks. Search for this
keyword in the title and you will be completely overwhelmed by
papers. What you may also notice is that these papers only started
appearing in numbers from about 1998 as figure
\fref{fcondmatpapers} shows.
Scan some of these papers and you will quickly notice that these
are not full of plots of conductivity and the like, but are
counting links on web pages \cite{AJB99} or discussing data on the
frequency of words in English texts \cite{CS,CS01b}. If you are
naturally skeptical about trendy new areas of physics and attempts
to mix physics with anything and everything, then the citations of
papers in journals of sociology \cite{Gran,PA93} and of books on
archeology and anthropology \cite{Brood,Davis,HH1,PH91} may just
be the last straw! However, one can not deny that an awful lot of
physicists have found something new and intriguing about networks
over the last few years.  What I hope to do in the article is to
give a flavour of what the excitement is all about, and perhaps
why underneath the hyperbolae accompanying any new developments,
there are sensible questions waiting to be answered.

First, what are these networks?  In the simplest form we are
discussing a collection of points or \emph{vertices} which are
connected by a variety of lines or \emph{edges}.  In more
complicated cases, we can add more information to our network. For
instance if it represents a transport network there may be
distances, times and/or capacities associated with our edges
(\emph{weighted} edges). Perhaps flow is only possible in one
direction along an edge so we think of them as having a direction
(\emph{directed} edges). Vertices may represent different types of
location, perhaps factories, warehouses and shops (\emph{coloured}
vertices). However, while such extra data is important in many
cases, networks are all fundamentally just a set of vertices and a
set of edges connecting some of the vertex pairs. Given that such
a simple concept is at the heart of many physical structures, it
should not be surprising that mathematicians have been studying
them for some time.  What I have been calling a network is what is
called a \emph{graph} by mathematicians and I will use the terms
network and graph interchangeably. Unfortunately, the terminology
of graph theory is not standardised so one must be careful to
check each author's definitions and to specify ones own notation.

Once a physical network has been represented as a mathematical
graph, then numerous problems can be considered. Optimisation
questions are often classic problems of graph theory.  One example
is the `travelling salesman' problem in which one must find the
shortest distance a salesman can travel given they are to visit a
set of cities (the vertices) along prescribed routes between
cities (the edges). Critical path analysis is another application
of graph theory where one tries to identify bottlenecks in a
process. However, the recent interest of physicists stems from the
discovery of new ways of classifying and generating graphs and
linking these new types to networks we see in the world around us.
Applications include various types of human interaction such as
social or business relationships. The internet is a very
fashionable and often overworked topic, but it is a natural
application for all that. In fact sociologists and anthropologists
had been studying some of these systems for some time, sometimes
using graph theory. However, physicists certainly provided new
tools and views even if the debate predated them.

So let me now turn to look at the different types of  network and
how they have been used to study mathematical and physical
problems.

% ************************************************************

\subsection*{Global Network Properties}

I will focus on the most basic properties of a network, and hence
I will ignore the any directions or weights associated with edges,
and any colours or labels added to vertices.  Further, I will
restrict myself to the case where multiple edges between the same
pair of vertices are not allowed, and an edge is not allowed to
start and end on the same vertex.  These networks are sometimes
called \emph{simple graphs} though the terminology varies. It will
also turn out that most networks of interest are \emph{sparse},
that is the number of edges $E$ is of the same order as the number
of vertices $N$.  This is considerably smaller than the potential
$N(N-1)/2$ edges of a simple network.

\begin{table}
\begin{center}
\begin{tabular}{r|l}
 Symbol & Meaning \\
 \hline
  $N$   & Number of Vertices in a network
\\
  $E$   & Number of Edges in a network
\\
  $k$   & Degree of a vertex
  \\
  $K$   & Average degree of vertices in a network
\\
 $n(k)$ & Number of vertices of degree $k$ in a network
\\
  $L$   & Average of the shortest distances between
  \\    & \hspace*{\fill} all vertex
  pairs in a network
\\
  $c$   & Clustering coefficient of a vertex
\end{tabular}
\end{center} \caption{Summary of various symbols used
throughout the text. See text for definitions.}
\end{table}

\begin{figure}[htb]
\begin{center}
  \scalebox{0.25}{\includegraphics{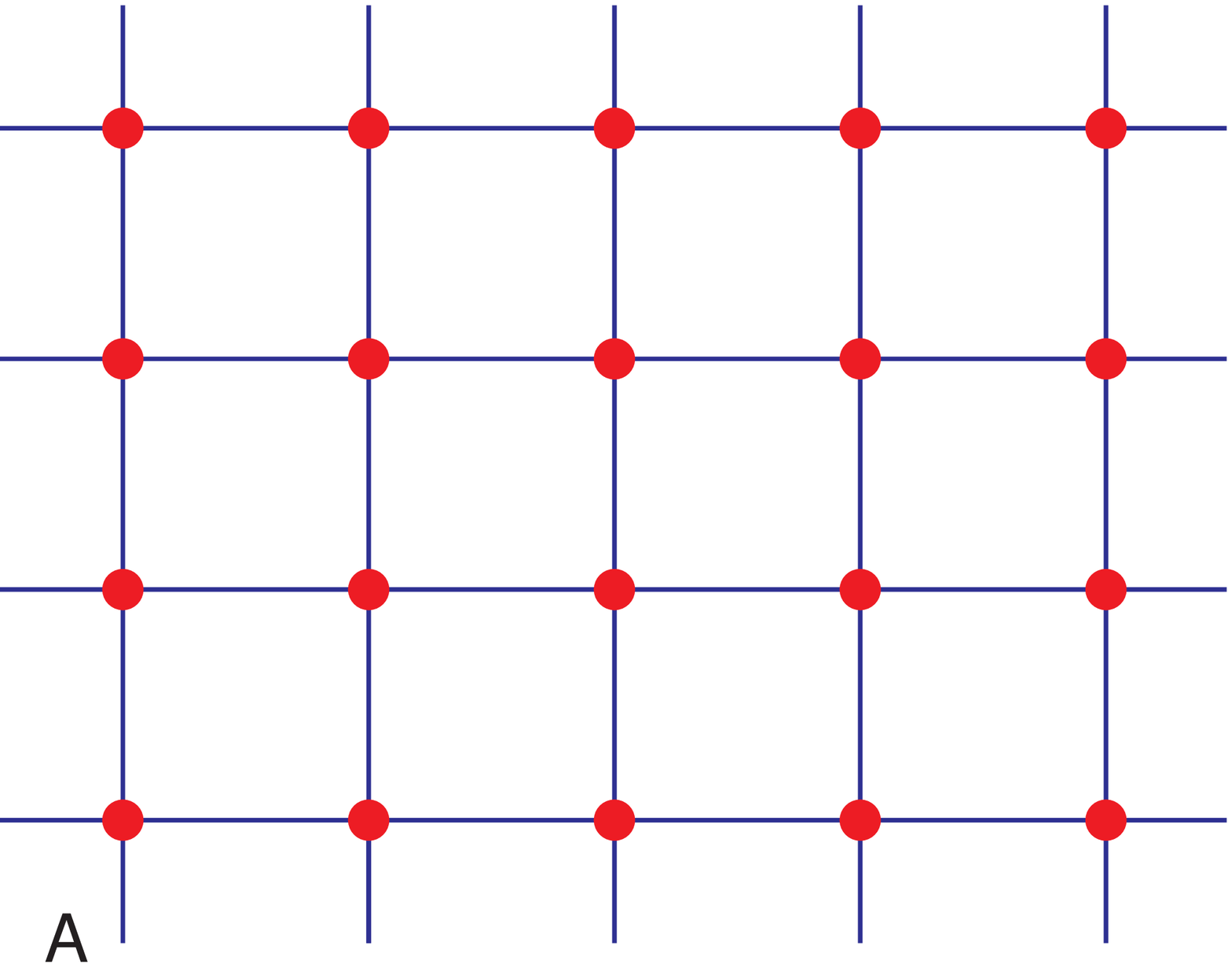}}
  \scalebox{0.25}{\includegraphics{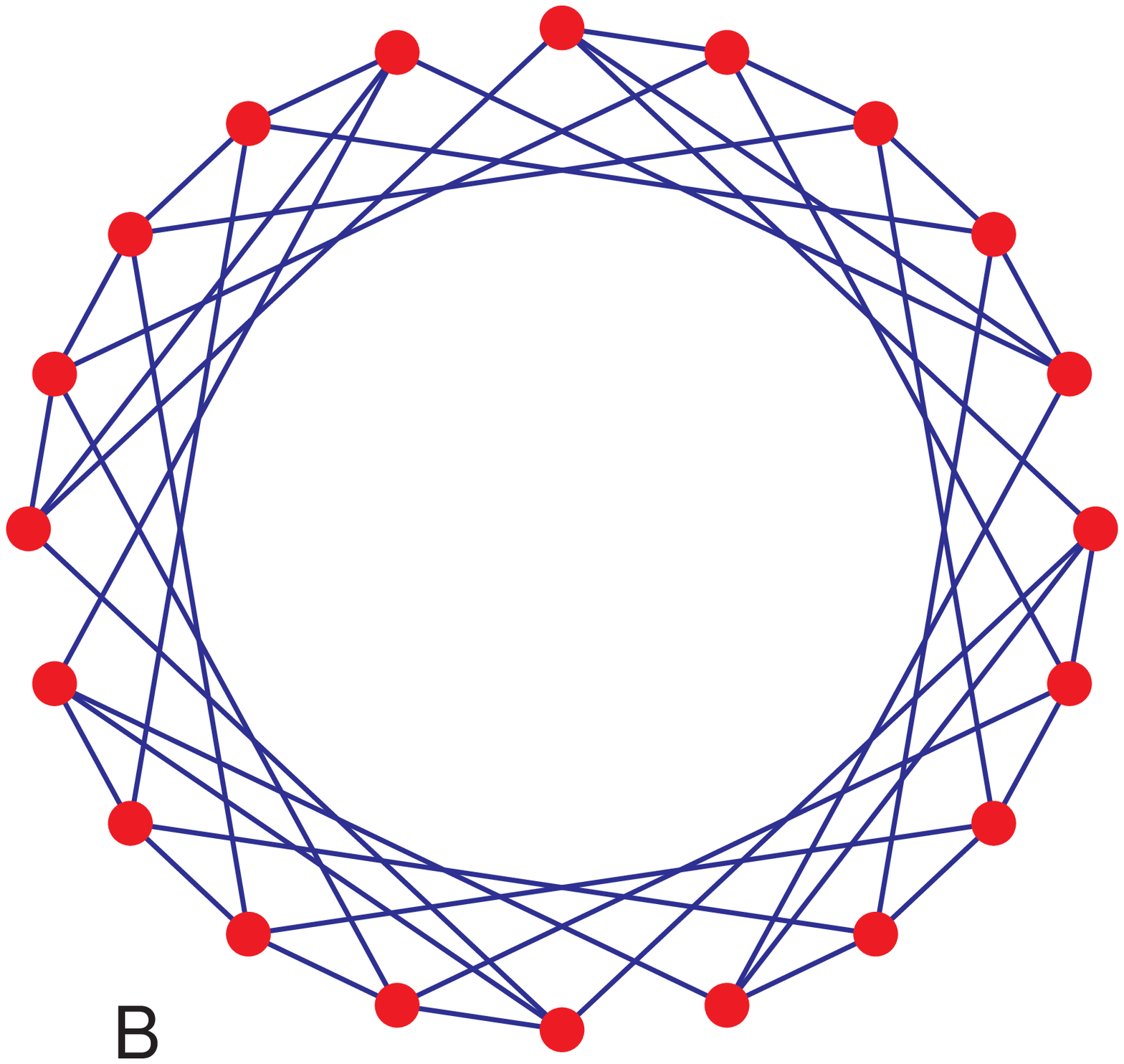}}
  \scalebox{0.25}{\includegraphics{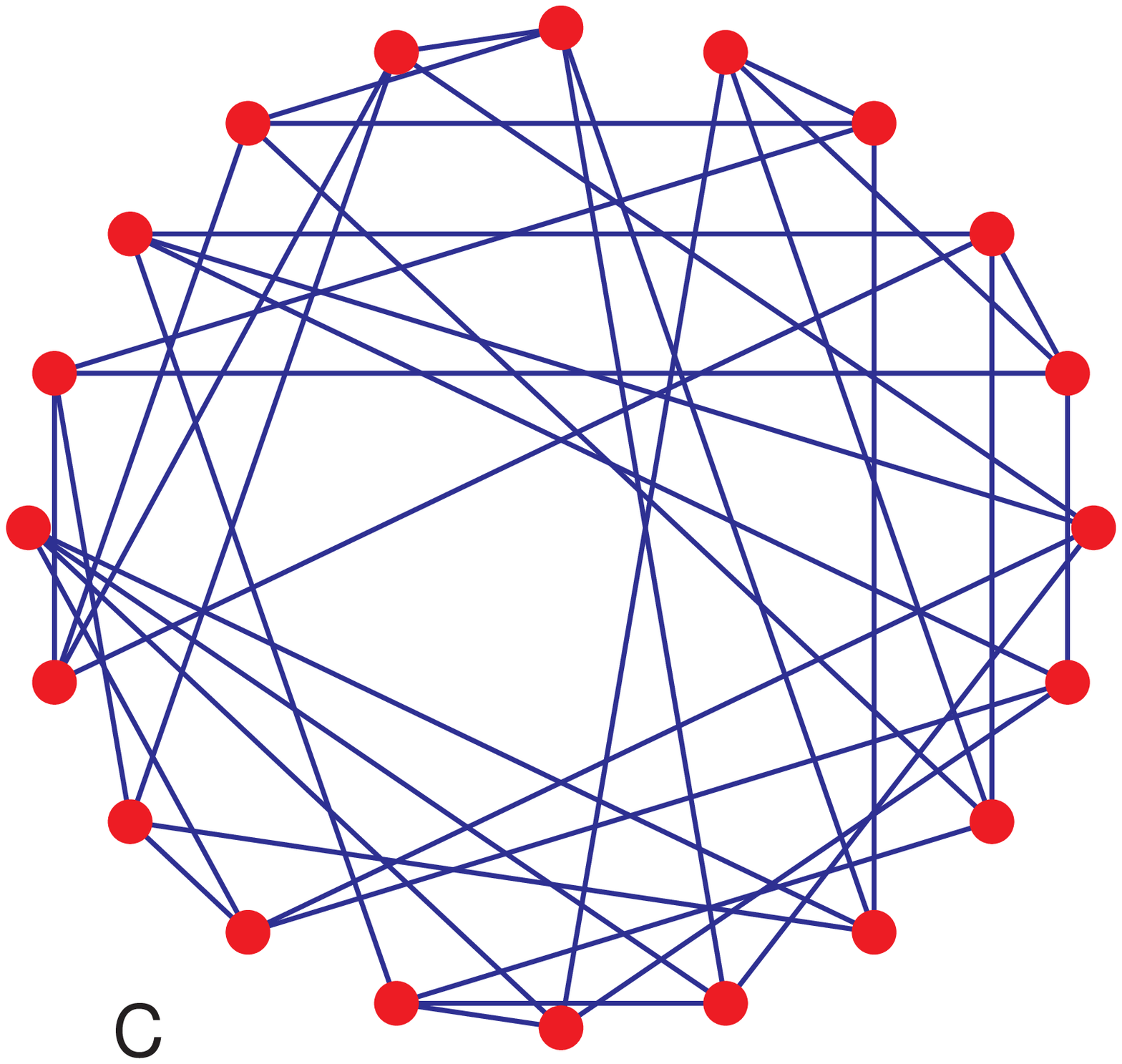}}
\end{center}
\caption{A 20 node square periodic lattice, with 40 edges between
nearest neighbours only, but shown in three different ways. In the
left hand picture A, edges going off the sides wrap around. In the
second two examples, the spatial coordinates associated with each
vertex are ignored. In the middle picture B, the vertices are
still displayed in a systematic order so the regular nature of the
lattice is still visible. In the last picture C the vertices are
displayed around the circle in a random order and the graph now
looks random, even though in terms of simple graphs it is as
regular as the other two.  For later reference note that this
lattice has an average distance of $L=2.32$, diameter 4, and the
average clustering coefficient is 0.}
 \label{flat2dn20per}
\end{figure}

As a physicist, the first type of network that comes to my mind is
the regular lattice.  Playing a fundamental role in solids, they
are characterised by invariance under translation in space by a
lattice spacing along a lattice axis\footnote{Strictly speaking,
only if they are infinite or periodic.}. In this case the vertices
of a network could represent the atoms of a crystal. The edges
could then indicate the most important interactions. For instance,
in a simple two dimensional regular square lattice, such as figure
\fref{flat2dn20per}, each vertex is attached to four edges so that
the \emph{degree} of every vertex is four to use the graph theory
terminology.  It therefore has $E=2N$ edges when there are $N$
vertices, and so it is a sparse network.

\begin{figure}[htb]
\begin{center}
%  \scalebox{0.25}{\includegraphics{Pajekn20k4rndcirc.eps}}
  \scalebox{0.25}{\includegraphics{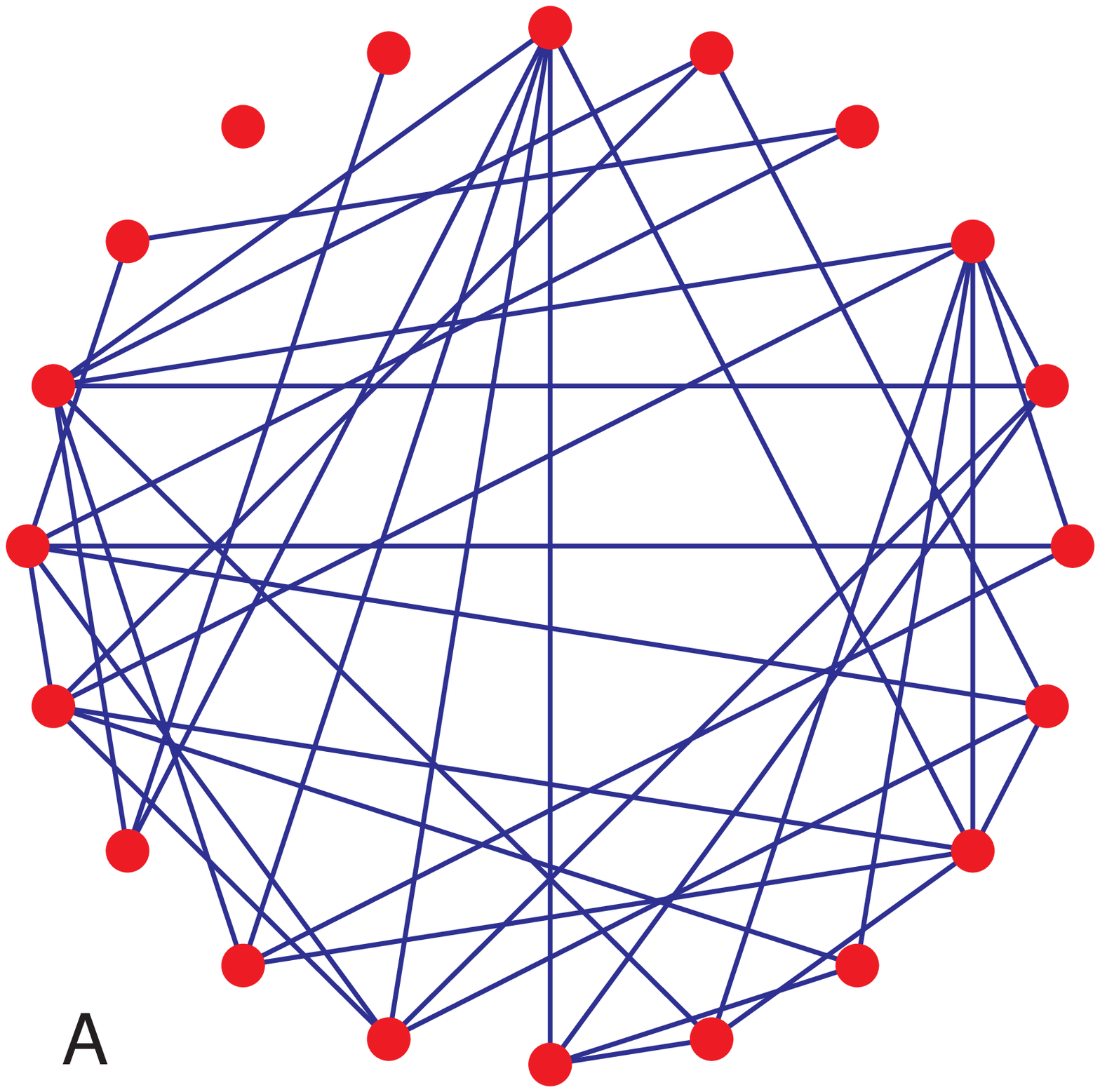}}
  \scalebox{0.25}{\includegraphics{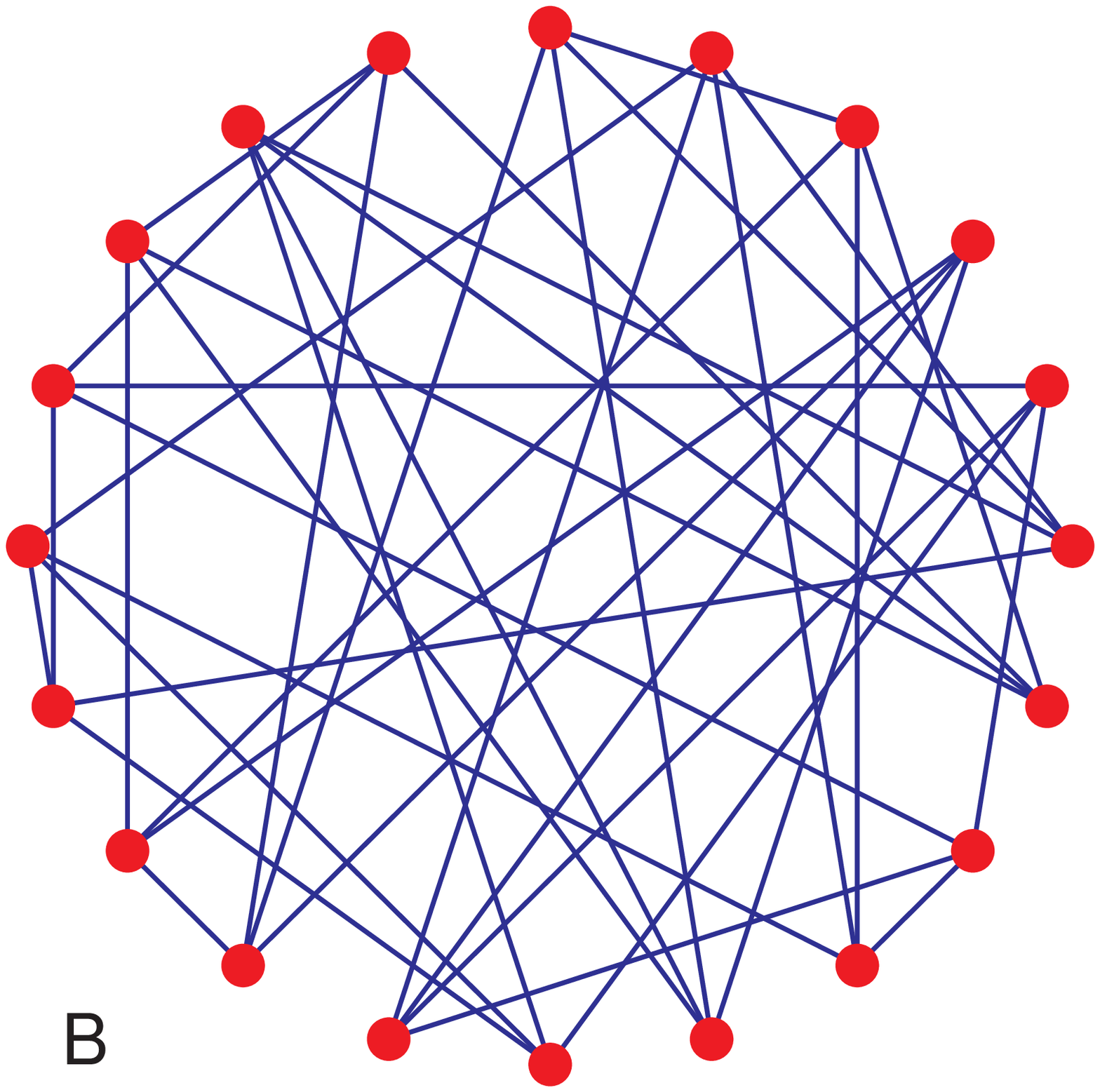}}
\end{center}
\caption{On the left a standard 20 node, 40 edge a random graph.
On the right a similar graph but constrained so that all vertices
are attached to four edges, i.e.\ their degree is always 4 like
the lattice earlier. For later reference, network A has average
distance $L=2.17$ (ignoring the disconnected vertex), diameter 5
and average clustering coefficient of $c=0.134$. Network B has
average shortest distance of $L=2.22$, diameter 4 and average
clustering coefficient of $c=0.15$.}
 \label{frndn20k4}
\end{figure}

In terms of abstract graph theory, it is easy to picture the
opposite of a regular lattice.  For instance if we took $N$
vertices we could select at random $E$ of the $N(N-1)/2$ different
vertex pairs and connect them to give what is called a
\emph{random graph}\footnote{Similarly, we could connect each edge
with probability $p=2E/(N(N-1))$ and on obtain $E$ edges on
average. I will confine the term random network to these
definitions though others sometimes refer to these as an
Erd\H{o}s-Reyn\'i, type random graph and extend the name to cover
a wider family of graphs formed by some probabilistic rule.}. If
$E$ is of the same order of magnitude as $N$ then we have a sparse
network again. For instance the random graph of the same number of
vertices and edges as the regular lattice of figure
\fref{flat2dn20per}, given in figure \fref{frndn20k4}, is sparse.
The random graph immediately highlights one point central in the
study of networks, namely that in general the vertices of a
network have no position associated with them. Clearly in the case
of the random network, the connections between vertices are made
without any reference to notions of relative position of vertices
unlike in the crystal example, and so we do not need to, and must
not, picture these vertices as having any position.  It is
therefore wrong to think of the random graph as living in any
particular dimension. Only the limitations of our visualisation
skills forces me to display the random graph of figure
\fref{frndn20k4} in two dimensions, and the implicit coordinates
of the vertices in the picture have no meaning. I could move them
anywhere on the page and, provided I maintained the edges
connecting the vertices, it would represent the same network.  The
central idea of graph theory and the issues studied with modern
networks is that it is the only the connections between vertices
that are fundamental and these and their associated properties
should form the starting point of any analysis.

Of course, the random graph of figure \tref{frndn20k4} has no
structure other than its connections, as this was all that was
used in its construction. The positions of vertices in some real
space are not an issue. However, the physical properties of a
material whose atoms are arranged in a square lattice do require
us to take account of the physical coordinates so in such a case
we must use the representation on the left of figure
\fref{flat2dn20per} and not those on the centre and right where
the positional information has been discarded to a greater or
lesser extent.

Problems with real crystals are not going to be advanced by
studying networks in themselves. However, there are many problems
where it is the connections themselves, and not their nature,
which are the key.  In this case the positions of the vertices in
our world are not important. Vertices could be people, the links
indicating a relationship which is not just a simple function of
the physical distance between people. The `space' of such
anthropological networks is not the simple Euclidean
three-dimensional one of most scientific problems. For instance,
suppose we require a system of radio communication between sites
(the vertices).  An edge linking two sites represents a
communication link and we will suppose that this requires a
dedicated frequency that no other pair of sites can use. Perhaps
for the system, the physical Euclidean distance between sites is
not relevant, being small enough not to effect error rates etc. On
the other hand, if the total bandwidth available is limited, then
we only have a limited number of frequencies available, we can
only have a certain number of edges in our network. If the error
rate in the communications rises with the number of edges
traversed by each message, then we need to find a network that
minimises the average number of edges traversed in moving from one
vertex to another on the graph, a purely graph based concept.

Let us now look at the various ideas and concepts that can be
defined with only the fundamental properties of a network.  In
other words, we can only work with a set of vertices and a set of
edges joining some of these vertex pairs. We can imagine walking
from vertex to vertex on this network, moving only along the edges
of the network.  In doing so we define paths on the networks
between vertices. In some networks there may be no path at all
between some vertex pairs. This means that the graph is
\emph{disconnected} and appears in two or more distinct pieces
called \emph{components}.
\begin{figure}[htb]
\begin{center}
  \scalebox{0.5}{\includegraphics{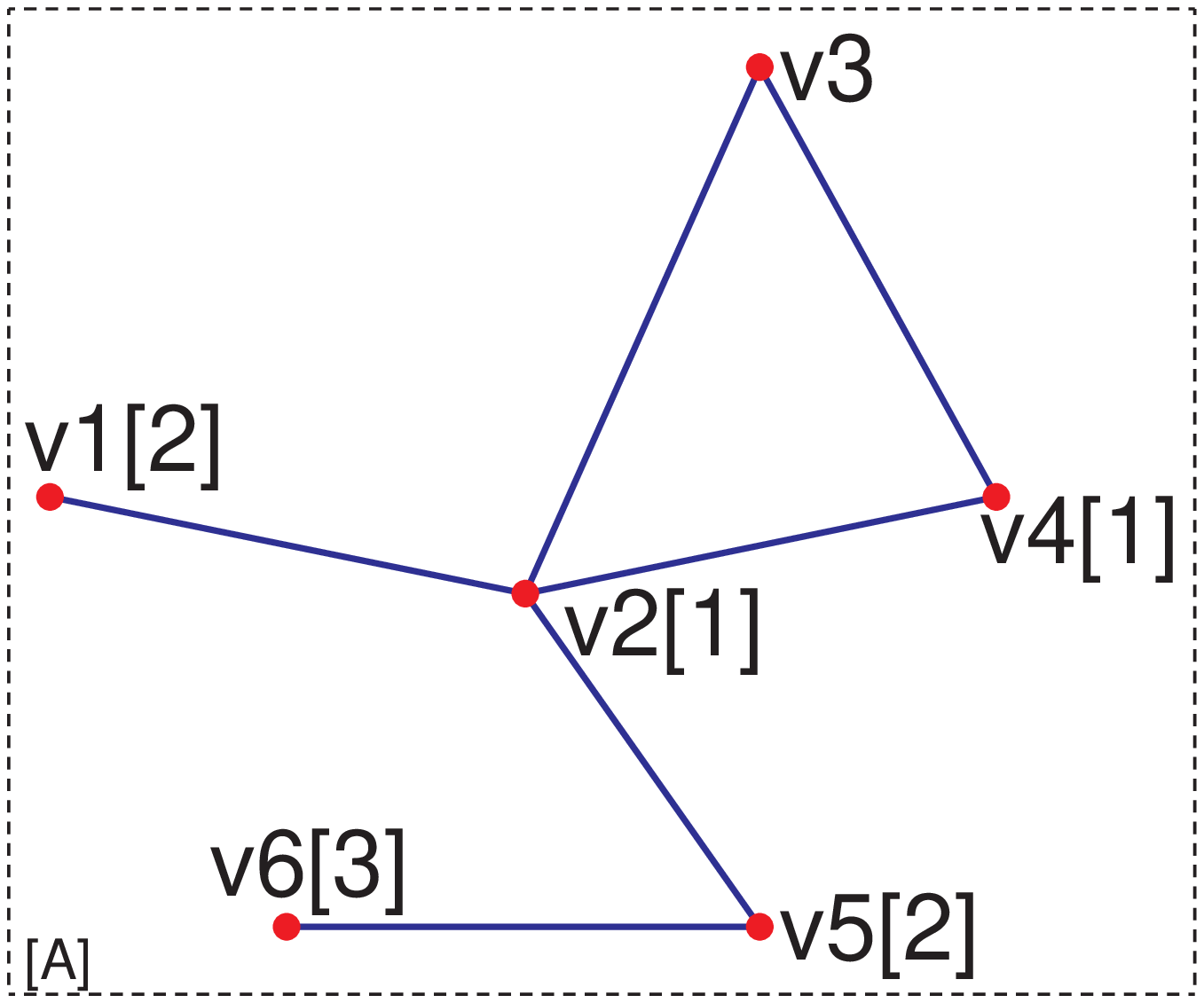}}
  \hspace*{0.5cm}
  \scalebox{0.5}{\includegraphics{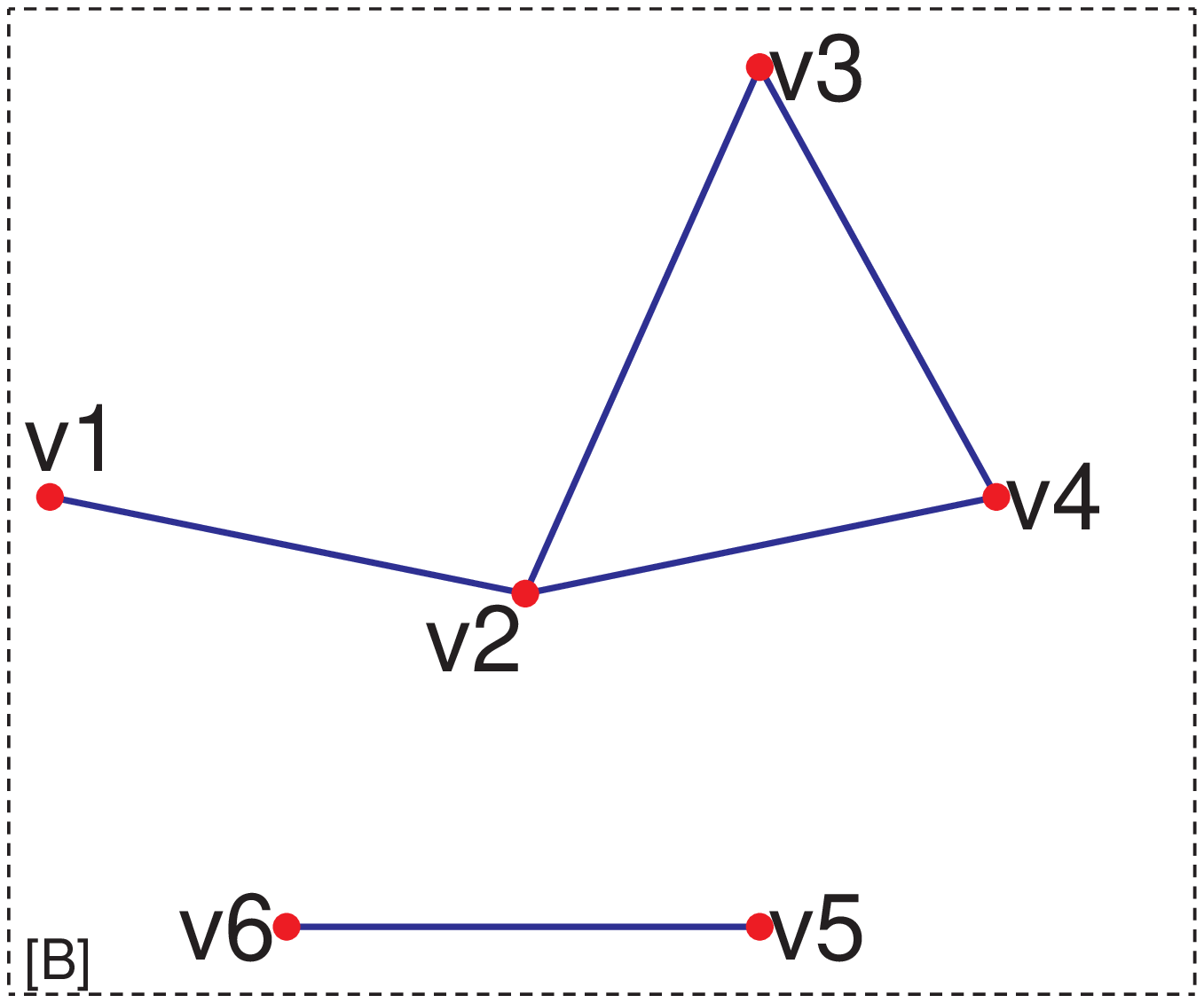}}
\end{center}
\caption{Graph [A] on the left is connected since one can walk
along the network from any vertex to any vertex.  The distances
from vertex v3 to any other vertex are given in square brackets.
So vertex v6 is 3 from v3, the length of the shortest path via v2
and v5 (paths via v4 for instance are longer and are not
considered). This is also the diameter of this network as no pair
of vertices has a shortest path longer than 3. The average of
shortest distances over all 30 pairs of vertices is $L=1.87$.
Removing the v2-v5 edge leaves the disconnected graph [B] on the
right since one can not then walk for instance from v1 to v6. The
v1,v2,v3,v4 vertices and their edges form one connected component
of the right hand graph.}
 \label{fsimpleex}
\end{figure}

Deciding if the graph is connected or disconnected is one of the
first things one should look at.  A regular lattice is clearly
always connected. On the other hand that is not the case for a
random network as we need a minimum of $(N-1)$ edges to connect
$N$ vertices (for instance in a one-dimensional lattice or in a
star formation with one vertex at the centre all others at the
edge).  Studying the connectedness properties of the random graphs
(for large $N$) was part of the seminal work done by the
mathematicians Erd\H{o}s and R\'enyi\footnote{Solomonoff and
Rapoport \cite{SR50,SR51} had introduced random graphs and proved
one classic result in 1951 but this did not seem to be well known
in the mathematical community.} in the late 1950's and early 60's
\cite{ER59}. Their work prompted the modern era of graph theory
development amongst mathematicians, as a look at one of the
standard texts by Bollob\'as \cite{Bollrnd} shows. What Erd\H{o}s
and Reyn\'i, showed was that a random graph is likely to be
connected only if at least $N\ln(N)/2$ edges are present for large
N. Put another way, if one increases the number of edges added
randomly to a graph, then there is a sudden change in the
connectedness property of the network.  This ought to make the
ears of a physicist prick up as it sounds like a phase transition.
Indeed the analogy can be carried through much further with
several characteristics of random graphs showing sudden changes as
we increase the number of edges, and one can use ideas used in
physics, such as percolation theory, to study these problems.

Let us continue to look at the basic ways of characterising
networks.  The paths between vertices along the edges of the
network have a natural length, simply the number of edges
traversed. Thus a natural measure of the \emph{distance} between
any two vertices is the length of the \emph{shortest} path between
the two vertices. If the graph is connected (or we restrict
ourselves to a connected piece or \emph{component}) then we can
define an average length scale $L$, that is the average of the
shortest distance between all $N(N-1)/2$ vertex pairs of the
network. A related idea is the \emph{diameter} of the graph which
is the largest distance between any two vertices in the graph.

For a regular lattice this network defined distance between two
points is called the \emph{Manhattan distance}, as we imagine that
this is the actual distance one has to travel between two places
in a city with a rigid grid street pattern.  For such lattices
embedded in real space there is also the usual Euclidean distance
between points $(0,0,\ldots,0)$ and $(x_1,x_2,\ldots,x_D)$, that
is $\surd (x_1^2 +x_2^2 + \ldots+x_D^2)$.  However, again we
emphasise that this is a special property of a lattice, namely
that it can be visualised as living in some $D$-dimensional space.
It is well to note though that, as the tourist in New York knows
and the example of the radio network showed, the network defined
distance and \emph{not} the usual Euclidean one can be the
relevant measure, even if the lattice network is physically
embedded in real space. For such lattice problems intuition based
on Euclidean distance experience can be misleading (see for
example \cite{BBDF}) and this again reminds us that we must leave
behind any pictures of our networks being embedded in some real
physical space.

So, how do a regular lattice and a random graph of the same number
of vertices and edges compare?  Roughly speaking the
size\footnote{As measured by diameter, average distances or other
similar distance measures.} of a D-dimensional lattice will grow
as $N^{1/D}$ while the size of the random graph grows much slower
as $\ln(N)$. In a large random network every node is much closer
to all the other nodes than in a comparable sized lattice. Thus,
for the earlier example of the radio communication network a
random network is better than a regular lattice.\tnote{However,
there are other global properties of a random network that may not
be so useful in a practical problem but which are intriguing for
theorists. Clearly we could connect $N$ vertices in a ring, a
one-dimensional lattice, with $N$ edges but it is unlikely that a
random graph of $N$ edges is connected. In fact at the critical
number of edges, $E_c \sim N \ln(N)/2$, where a random graph
becomes connected, the distance between nodes, judged either by
average distance or diameter, is at a maximum for a random graph.
When the graph is disconnected one often adopts the distances and
diameters within connected parts as a measure of distance,
ignoring the fact these measures are not defined for the whole
disconnected graph.  It is these distance measures which reach a
maximum as $E$ rises towards $E_c$ from below in a random graph.
This is \emph{exactly} the behaviour seen at a second order phase
transition where there are correlations over scales as large as
the sample.}

\subsection*{Local vertex properties}

So far ideas such as connectedness, average distance and diameter
are reflecting the properties of the network on a large scale. For
many problems the local neighbourhood of vertices may play a vital
role.  This is familiar to physicists where the coordination
number of a lattice --- the number of nearest neighbours --- is
important in many problems and provides a simple way of
highlighting the differences between say a triangular and a square
lattice. In graph theory neighbours are defined only by the edges
of a vertex so the number of nearest neighbours is the number of
edges attached to a vertex, and this is called the \emph{degree},
$k$, of a vertex. The average degree $K$ is simply $K=2E/N$ where
$E$ and $N$ are the numbers of edges and vertices respectively.
Thus for a sparse graph this number should be of order one, or at
least $K$ must not grow as fast as $N$ as we increase the size of
a network. For simple lattices all vertices have the same degree:
in $D$ dimensions it is $K=D(D+1)$ for a hyper-tetrahedral
(equilateral triangle faced) lattice, $2D$ for a hyper-cubic
(square faced) lattice. However most other networks, such as the
random graph, have vertices of a wider variety of degrees.  This
is best summarised by giving the \emph{degree distribution}
$n(k)$, the number of vertices in the network with degree $k$.
This is a delta function for a lattice, but for many other graphs
it is a distribution with a tail.

\begin{figure}[htb]
\begin{center}
  \scalebox{0.8}{\includegraphics{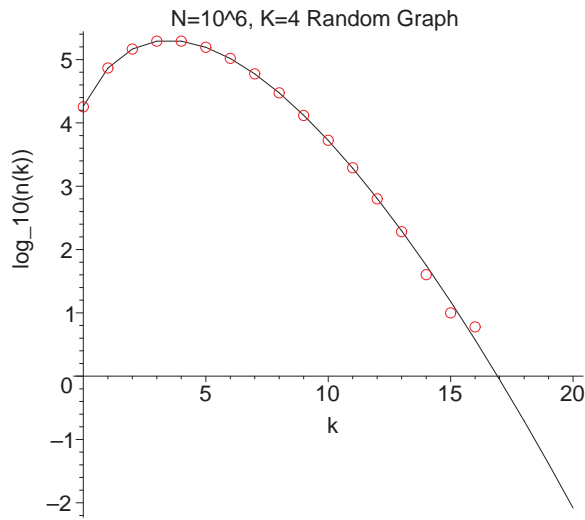}}
\end{center}
\caption{Plot of degree distribution $\log_{10}(n(k))$ against
degree $k$ for random networks of with $N=10^6$ vertices and
average degree $K=4$. The line is the most likely distribution for
a random network, a binomial distribution. Circles are for a
single example of a random network generated by computer and in
this case 6 vertices had the largest degree of 16. The probability
of finding a node with degree above twenty falls off as
$\exp(-k\ln(k))$.}
 \label{frndn1e6k4}
\end{figure}

For the random graph it is simple to show that the degree
distribution is binomial with $n_\mathrm{rand}(k) = N
p^k(1-p)^{(N-1-k)} \Csymbol{N-1}{k}$ where $p$ is the probability
that any given pair of vertices is connected. On average there are
a total of $E=pN(N-1)/2$ edges with most vertices having a degree
within\tnote{$\sigma^2 = K(1-K/(N-1))$} $3\sqrt{K}$ of the average
degree $K=p(N-1)$. For large $N$ and fixed $K$, this is to a good
approximation a Poisson distribution $n(k) \approx N K^k
\exp(-K)/k!$ whose tail falls off slightly faster than an
exponential, roughly as $n(k) \propto \exp ( -k\ln(k))=k^{-k}$.
The fast fall off means that there are essentially no vertices
with degree bigger than $k_c$ where $n(k_c)=1$ . Solving we find
that\footnote{A more accurate formula is $k_c(\ln(k_c) -1)=
\ln(N)-K$.} $k_c\ln(k_c) \approx \ln(N)$ which is in most
practical cases close to the most likely degree for a vertex $K$.
For instance a random graph with $K=4$ and $N=10^6$ will probably
have no vertex with degree larger than 17, much smaller than the
potential $(10^6-1)$ edges available. This is clear from the
distribution shown in figure \fref{frndn1e6k4}.

\subsubsection*{Clustering}

While the degree distributions show clear differences between
lattice and random networks, there is another measure of their
differences which captures some of the order present in the
neighbourhood of a lattice vertex but missing in a random graph.
This is the \emph{clustering coefficient} $c$. In the simplest
form we pick a vertex and look at all of its neighbours, the $k$
vertices connected to it by an edge of the network. We then see
what fraction of the $k(k-1)/2$ possible neighbour-neighbour edges
are present in the graph and this is $c$.
\begin{figure}[htb]
\begin{center}
  \scalebox{0.5}{\includegraphics{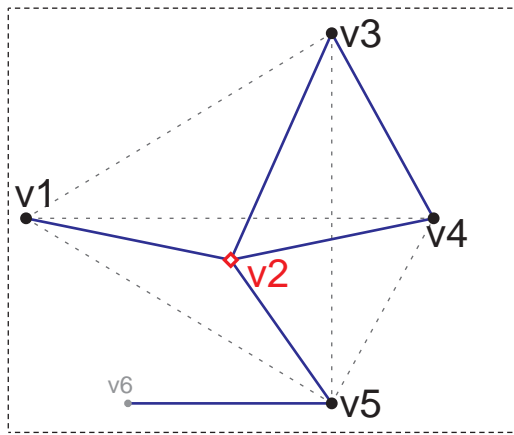}}
\end{center}
\caption{The simplest cluster coefficient illustrated for the
diamond vertex v2.  It has four nearest neighbours, v1, v3, v4 and
v5.  Of the 6 possible edges between the neighbours of v2, only
one, the v3-v4 edge is present.  The other potential
neighbour-neighbour edges, indicated by dashed lines, are not
present. Thus the clustering coefficient of v2 is $c=1/6$. The
vertex v6 plays no role in this calculation. The clustering
coefficients of the vertices  v3 and v4 are $c=1$ while for v5 it
is $c=0$.  The clustering coefficient is not defined for vertices
with only one neighbour such as $v1$ and $v6$ here.}
 \label{fsimpleclustercoef}
\end{figure}
On a regular triangular lattice in 2d one would have $c=6/15=0.4$,
reflecting its close knit local structure.  Of course a square
lattice has $c=0$ and to see its tight local structure we would
have to generalise our definition to $c_2$ involving neighbours
and next-to-nearest neighbours\footnote{We could look at the set
$N_2$ of all nearest neighbours and next-to-nearest neighbours,
and define $c_2$ to be the fraction of possible edges between
these vertices which are actually present in the graph. So on a
square lattice in two-dimensions every vertex has 12 vertices in
its set $N_2$ with 55 possible links edges them of which only 12
are present so $c_2=12/55 \approx 0.22$}.  However, in most
networks the clustering coefficient $c$ varies from vertex to
vertex and an average is usually quoted\footnote{As the clustering
coefficient $c$ is not defined for vertices with a degree less
than two, the averages quoted in this article are taken only over
vertices where $c$ is defined.  There is a second weighted average
which is useful and often encountered in the literature but
unfortunately the two definitions are not always clearly
distinguished \cite{BR02}.}.

In some problems local communication between neighbouring vertices
is essential. In lattice Monte Carlo simulations, the algorithm
requires a large amount of information to be shared between
neighbouring lattice sites, reflecting the local spatial (and
possibly temporal) nature of interactions in many problems, such
as atomic spins of a crystalline material. Thus in a parallel
computer built for this problem, the communication network linking
the processors should have a large amount of local structure, and
indeed a lattice configuration is a common solution.  On the other
hand, the internet today, and computing networks such as the Grid,
are designed for a great range of tasks which require different
computing/data centres (the vertices) to be able to communicate
easily with most other sites at different times. Local structure
is not paramount. The human brain faces similar issues. The
translation of the sound of a spoken command into an action
involves processing in several different areas of the brain, each
area specialised and involving many neurons, yet each area must
communicate with many others.

The random network lies at the other extreme from a lattice, and
has the minimum local structure. In a random network there is no
preference for neighbours to connect to each other rather than
anyone else.  In other words, all edges are treated equally and
they are present with probability $p$, and not present with
probability $(1-p)$.  In a lattice, neighbours of neighbours of
some chosen vertex $i$ can not be very far away in terms of
Euclidean distance\footnote{We'll assume for simplicity that our
lattice exists in real space as a spatial lattice.}, no more than
two lattice spacings. This means that for a lattice we are
guaranteed that the neighbours of neighbours are also going to be
close to the original vertex $i$ in a Euclidean distance sense,
and so may well be connected to them. They will have a far higher
probability than a random chance $2E/(N(N-1))$ of being connected.
In a lattice, vertices are clustering together in a network sense.
On the other hand, in a random graph there no reason why the
neighbour of a neighbour of the original vertex $i$ has any other
relation to $i$. There is no space we can use to embed our network
in, no Euclidean distance argument to use. Thus we should expect
that a random network should have a much lower clustering
coefficient than a similar sized lattice, reflecting the fact that
we can always rearrange a lattice network so it looks like the
usual regular arrangement in real space so all neighbours are
close to each other in all senses.\footnote{It does not matter if
our lattice network really exists as a lattice in a real space.
The network properties are the same whether we look at the right
hand or left hand version of figure \fref{f1dlatrw}.  Its just
much easier to visualise the local network properties of a lattice
if we exploit the picture on the left.}  Indeed a quick
calculation shows that a random graph of $N$ vertices, has
$N(N-1)/2$ possible edges but if only $E$ edges are present, then
on average a vertex will have a clustering coefficient of
$c=2E/(N(N-1))= K/(N-1)=p$, just the probability $p$ that any one
potential edge in a random graph is present. For a sparse random
graph of a reasonable size, this is much less than 1. Indeed,
while clustering coefficients are constant for a lattice as it is
made larger, for a random graph of the same number of vertices,
$N$ and same number of edges $E=KN/2$ as the lattice, since $K$ is
fixed for a lattice, the clustering coefficients drop as $1/N$.

\subsubsection*{Small Worlds}

We have seen that large sparse random graphs have as little local
network structure as is possible for a network since all edges are
treated equally so there is no special treatment for neighbours.
However this also means that the distance is small.  All vertices
are treated equally so each new vertex you visit on a path is
likely to throw open $K$ new vertices unreachable until now (at
least for the first few steps on a route). A random graph has a
small diameter.  On the other hand, the way a lattice of the same
size uses its edges to provide local structure and this means that
it takes a long time to reach an arbitrary vertex, and so a
lattice has a large size in network terms.  These differences are
exacerbated as we consider networks with more and more vertices.
On the lattice, the average degree $K$ is fixed, as the local
structure is constant by definition, but the size of the network
grows as a fractional power $N^{1/D}$ where $N$ is the number of
vertices and $D$ is the dimension. In the corresponding random
network of the same number of edges and vertices (fixed $K$ and
growing $N$)  the probability $p=K/(N-1)$ of any one link being
present is dropping so the clustering coefficients tend to zero.
However, the distances rise only as $\ln(N)$, much slower than any
lattice.

This poses a question.  Is there a sparse network which for some
given number of vertices $N$ and edges $E$ has the local structure
of the lattice but the small network size of the random graph?
This is particularly important when we think about human networks.
In 1967 Milgram reported on an experiment in which he asked people
in Omaha, Nebraska and Wichita, Kansas to send packets to one of
two people in Cambridge, Massachusetts specified by name,
profession and rough location only.  However, packets were to be
passed only between people who know each other by first name.  In
this way, Milgram hoped to map out the social network of close
friends.  The result was that if the packets arrived at the
correct person, they had been through about five intermediaries.
This is much smaller than we might have guessed given the physical
distance between the original senders and the final recipients,
who also presumably had no direct social or other contacts.  This
is the idea of the \emph{small world}, namely that though we may
think of ourselves as living amongst a small group of friends and
colleagues, we all have a few contacts outside any circle and
through these friends, and their friends of friends we are no more
than a few handshakes from any person in the world.  In terms of
networks, vertices are people and edges represent links between
people who are on first name terms. Milgram's experiment showed
that the average distance between sources and targets, if packet
were delivered successfully, was no more than six. As Guare puts
it in his 1990 play
\begin{quote}
``I read somewhere that everybody on this planet is separated by
only six other people. Six degrees of separation. Between us and
everybody else on this planet. The president of the United States.
A gondolier in
Venice.'' ... \\
 ``It's not just the big names. It's anyone. A native in a
rain forest. A Tierra del Fuegan. An Eskimo. I am bound to
everyone on this planet by a trail of six people. It's a profound
thought.'' ... \\
 ``How every person is a new door, opening to other
worlds. Six degrees of separation between me and everyone else on
this planet.''
\end{quote}

In fact this type of behaviour is common in many areas of human
interaction so references to it are not uncommon, and certainly
predate these examples.\footnote{Barab\'asi \cite{Bar} mentions an
obscure short story ``L\'ancszemek'' (Chains), paradoxically by a
famous Hungarian author Frigyes Karinthy published in 1929 where
it was suggested that it took five acquaintances to reach anyone.}
I remember some excited students showing me the ``Kevin Bacon
game'' and its web site \cite{OoB} for the first time. Name an
actor, and the web site will tell you their `Bacon number', the
number of steps it takes to reach Kevin Bacon via actors who are
paired if they appeared in the same film. For instance Charlie
Chaplin has a Bacon number of 3 since he appeared in ``Screen
Snapshots: Spike Jones in Hollywood'' (1953) with Douglas
Fairbanks Jr., who in turn was in ``Hollywood Uncensored'' (1987)
with Eli Wallach and Mr Wallach was in ``Mystic River'' (2003)
with Kevin Bacon.  The actors are vertices, edges represent a film
in which both the actors associated with ends of the edge
appeared. This amusement shows that paths between actors are
surprisingly short, around 3.7 on average (in 1999 \cite{Watts99})
with lower numbers for actors such as Kevin Bacon who at the time
of writing was only 2.944 from another actor on
average\footnote{The web site states that on 29th April, 2003 Rod
Steiger had the shortest average distance, average Bacon number of
2.652, so disproving the theory behind the student bar game that
Kevin Bacon was the centre of the Hollywood universe. Mr Steiger
was just ahead of Christopher Lee at 2.660 while Mr Bacon's
average was 2.941 putting him only 1222nd on the list on that
date. The network is constantly growing and at the time of
writing, February 2004, the largest connected component had
$6.2\times 10^5$ vertices (actors) with an average degree of
$K=2.94$.}. However, long before computers and the internet
enabled this game to be played with ease, the mathematicians had
the same concept of an Erd\H{o}s number, the same Erd\H{o}s that
produced the seminal random graph papers. The vertices were
mathematicians who were linked if they had coauthored a
paper.\footnote{Both the academic coauthorship and Kevin Bacon
game examples can be played in a different way. We could have the
films (academic papers) as vertices with the actors (academics) as
links.  Get from ``Ben Hur'' (1926) to ``Ben Hur'' (1959) (from
\tpaptitle{On Random Graphs I} \cite{ER59} to
\tarttitle{Collective dynamics of `small-world' networks}
\cite{WS}) in as few steps as possible. This illustrates the way
that the same data can often be represented as a network in many
different ways.} Erd\H{o}s coauthored with a large number of
people, a result of a fabled itinerant lifestyle, an innate
mathematical ability of his own and apparently an ability to
encourage and stimulate the work of others.  The mathematician's
game is a tribute to Erd\H{o}s \cite{ErdNo}.

The underlying idea in all these examples is that these networks
reflect human social interactions. They have a lot of local
structure, many of our friends are our friends' friends too, many
actors appear in the same type of films which therefore draw on a
small pool of actors active in that genre at the time, academics
often collaborate several times with the same people. In terms of
networks, they have a relatively large clustering coefficient. How
large is `large'?  Well the best way to put it is that it is much
larger than a random graph of an equivalent size. On the other
hand, the distances across the network are small, comparable with
those obtainable from a random graph and much smaller than any
regular network.  This then is the definition of a small world
network\footnote{Well, except that some people drop the clustering
part of the definition.  You just can't win with the nomenclature
in this field.}, namely $c_\mathrm{sw} \gg c_\mathrm{random}$,
$L_\mathrm{sw} \sim L_\mathrm{random}$. This is all very well, but
what type of model gives us a small world?

\begin{figure}[htb]
\begin{center}
  \scalebox{0.22}{\includegraphics{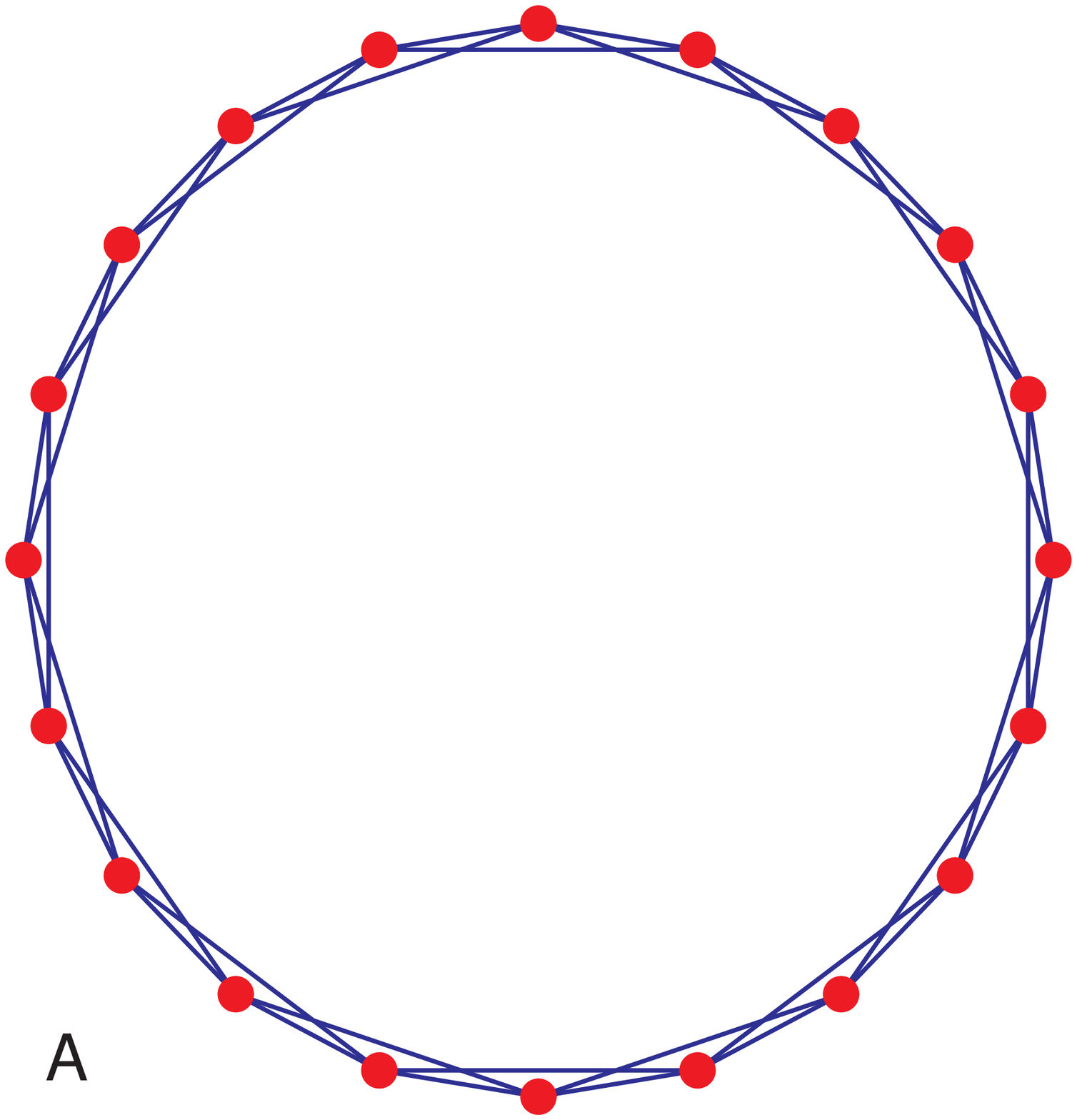}}
  \scalebox{0.22}{\includegraphics{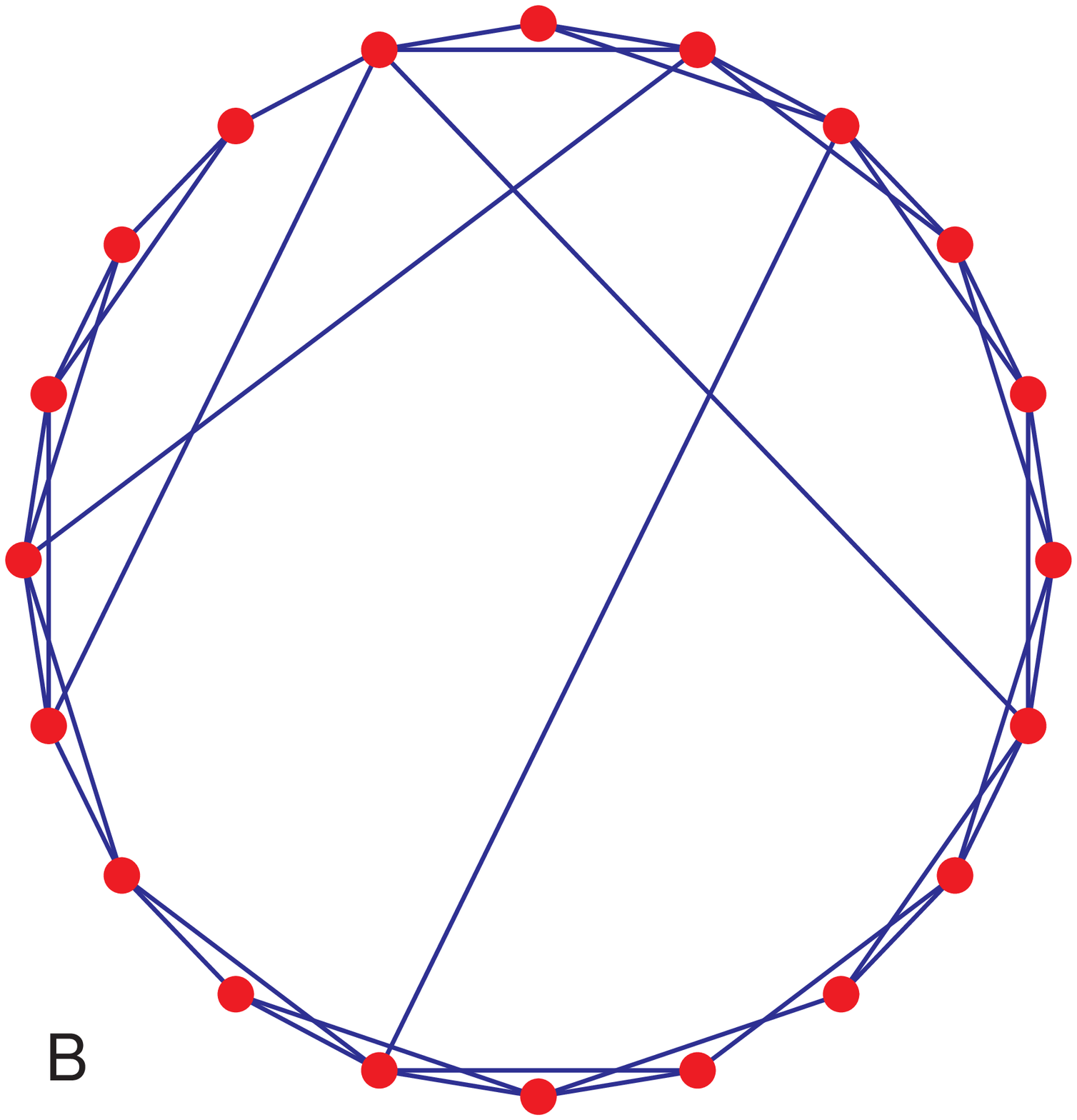}}
  \scalebox{0.22}{\includegraphics{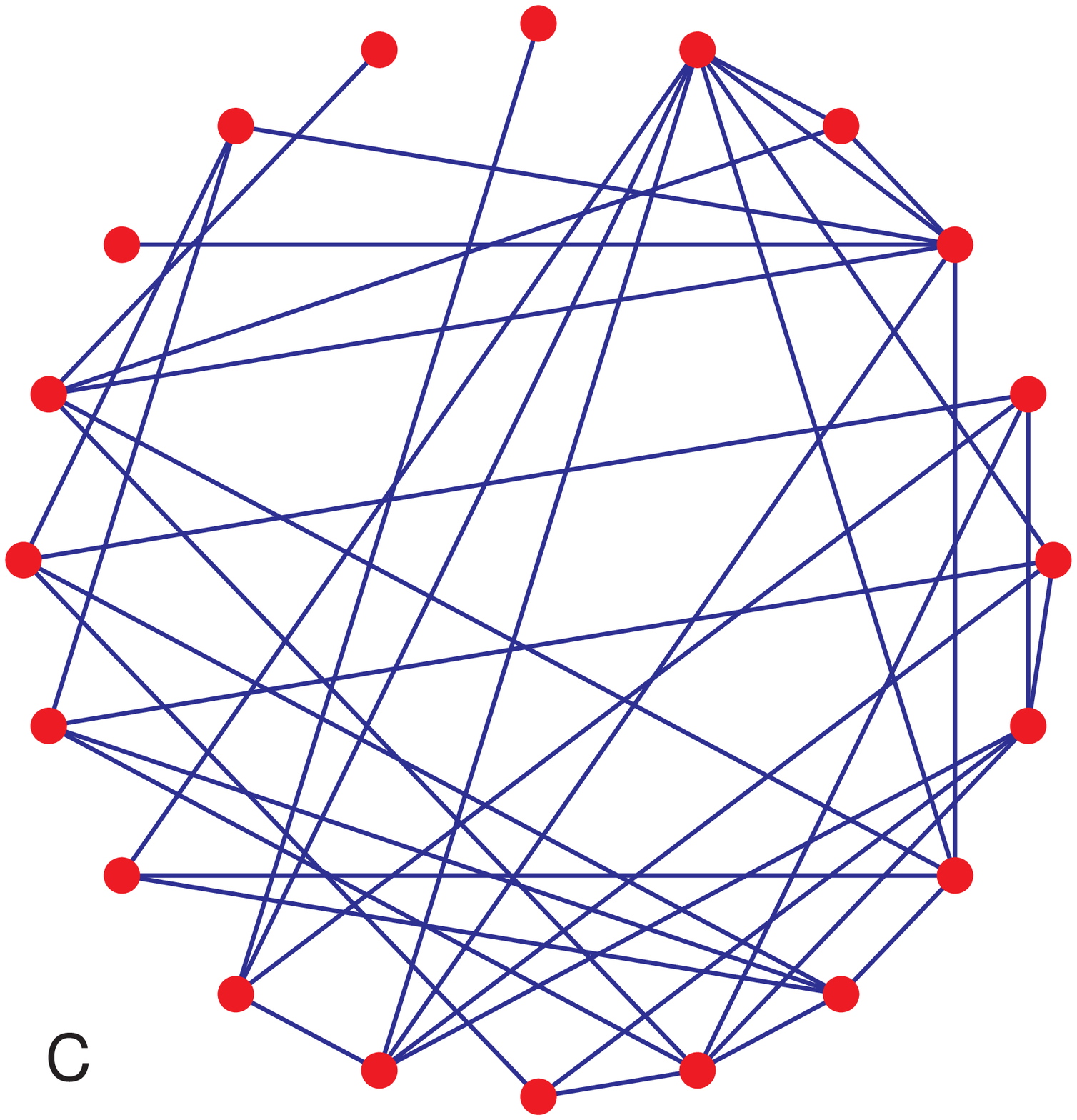}}
\end{center}
\caption{Evolution of a 20 node, 40 edge network through rewiring.
On the left, network A is a regular one-dimensional lattice with
neighbours and next-to-nearest neighbours connected by edges.  In
the middle, B is the same graph with 5 randomly chosen edges
reassigned, and finally on the right, C is the same graph after
200 such rewirings. The order of the vertices around the circle is
the same in all cases and no more than one edge between any vertex
was allowed. From left to right the average distance are
$L=2.89,2.35,2.21$, average clustering coefficient are $c=0.5,
0.40, 0.23$ and diameters are 5,4,5.}
 \label{f1dlatrw}
\end{figure}

\begin{figure}[htb]
\begin{center}
  \scalebox{0.7}{\includegraphics{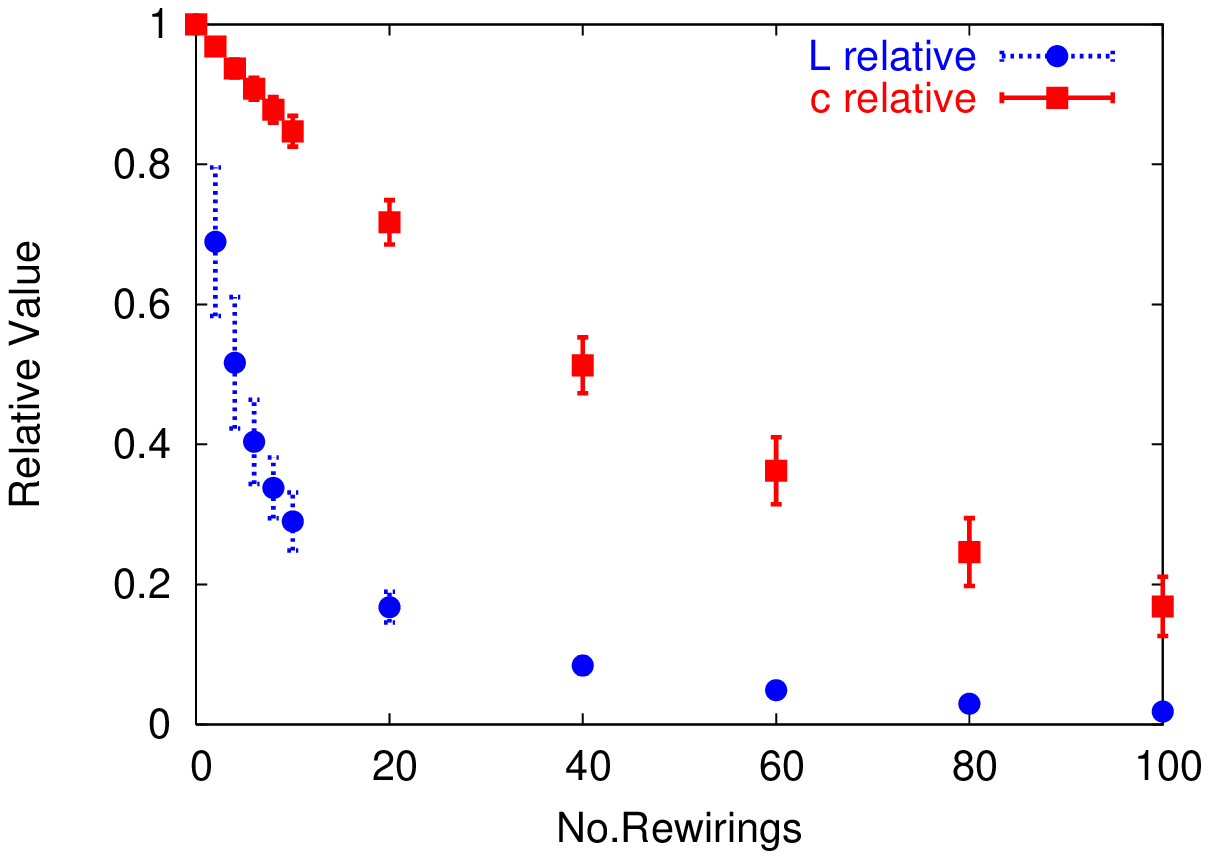}}
\end{center}
\caption{Plot of $c_\mathrm{rel}$ (top red square points) and
$L_\mathrm{rel}$ (bottom blue circle points) against the number of
rewirings (proportional to $p$ for small $p$) in an WS rewiring
scheme for 100 runs starting from a 100 vertex $K=4$
one-dimensional lattice. The value $c_\mathrm{rel}=
(c-c_\mathrm{rnd})/(c_\mathrm{lat}-c_\mathrm{rnd})$ is the
clustering coefficient scaled so that it is 1.0 for the lattice
(left hand side) and 0.0 for the random graph (right hand side).
$L_\mathrm{rel}$ is scaled in a similar manner.}
 \label{fwsrw}
\end{figure}

This is where the seminal paper of Watts and Strogatz of 1998
comes in \cite{WS}.  Their idea was to start from a lattice, in
their case a one dimensional ring with neighbours and
next-to-nearest neighbours linked, so the average degree was
$K=4$. You then look at each of the $E=2N$ edges in turn and with
probability $p$, you \emph{rewire} this edge, that is you choose
two new vertices at random and place the edge between
them.\footnote{In fact their algorithm was slightly different in
implementation and there are many slightly different algorithms
which achieve the same result. The key idea is adding a few random
links to a regular lattice gives a small-world network while
adding more random links eventually brings you to a random
network.} This is illustrated in figure \fref{f1dlatrw} in what is
perhaps the iconic image of this topic in its modern era.
If $p=1$ every edge is placed at random and we have a random
network of the same number of vertices and edges, short distances
but little local structure. If $p=0$ we have the original lattice,
high on local structure but with large distances.  For
intermediate $p$ we might expect a hybrid of these features with
clustering coefficient and distance to drop together as we
increase $p$. In fact Watts and Strogatz highlighted the fact that
as $p$ was increased, the distances dropped rapidly down to random
network levels, while the clustering coefficient only came down
slowly at first (linearly for small $p$). Thus for a small range
of $p$ values in their model, they were producing a small world
network retaining most of the local structure of the lattice but
having the short distance characteristic of the random graph.
Results for an $N=100$, $K=4$ case are shown in figure
\fref{fwsrw}.
These are easy to understand qualitatively.  Rewiring one or two
edges will only lower the clustering coefficient of a few
vertices. For the first few rewirings then, the average $c$ is
going to come down in proportion to the number of rewirings.  On
the other hand, while removing an edge makes little difference to
most paths between vertices, putting it back is likely to create a
shortcut between distant vertices on the lattice. Many shortest
paths between vertices will be now be shorter as many paths can
take this one shortcut. The average distance drops dramatically
for the first few shortcuts.

As we've noted, the idea of a small world network was not new. The
WS (Watts and Strogatz) model is just one exemplary model of a
small world but it offered the prospect of simple models of new
types of network and a systematic way of characterising them.
This, when coupled to modern desktop computing power, allowed any
scientist to make a systematic and statistical study of networks.
As in the case of the random networks, analytical results are hard
to come by so the ability to do numerical experiments is central
to this field. The same computing power, along with the ability to
gather and exchange large amounts of data, also enabled the
analysis of real data along the lines suggested by the work of
Watts and Strogatz. Of course, others, notably the social
scientists, had been analysing such data using computers and the
language and techniques of graph theory for some time. However,
Watts and Strogatz opened the doors for physicists to join in as
the explosion of preprints on networks placed on the archive
\texttt{cond-mat} shows.   In 2003 the number of papers containing
a word starting with ``network'' in their title was 730\% higher
than in 1997, see figure \fref{fcondmatpapers}.

\subsubsection*{Hubs}

Not surprisingly a random solution is not a good solution for the
radio communication problem suggested above. One solution which
can connect more sites for the same number of edges is where one
site is connected to every other site, all of which have only that
one edge. What we then have is a single dominant hub in our
network and all paths in the network will go through this hub.
This means that all vertices are relatively close to each other in
terms of the natural network distance, as the path from each
vertex to the hub is relatively short and then again from hub to
any destination vertex takes only a few edges. While short
distances are a feature of random networks and small world
networks, this is not the whole story. The random network and the
WS small world network do not have great big hubs.  There the
exponential nature of their degree distributions $n(k)$ for larger
$k$ means that the vertices with the most connections are of a
relatively small size in a sparse network. As noted above and
shown in figure \fref{frndn1e6k4}, in a random network of $N=10^6$
vertices and average degree $K=4$ there is unlikely to be a vertex
with degree higher than 17. Thus these types of network do not
allow for very big hubs, such as we might expect to see in some
real world situations or with the star formation solution to the
radio network problem.   Of course, there are practical
limitations on hubs in many problems which prevents such extreme
hubs, as the centre of a star formation, appearing. For instance
in communication networks dealing with too many channels
simultaneously at any one point is likely to be impossible or
unrealistically expensive. Still, it seems that for some problems
large hubs are preferable and it should not be too surprising to
find that there are many studies which seem to see many more large
hubs than one would expect from random or WS small world networks.

In order to have hubs, we need a degree distribution $n(k)$ with a
much longer tail than the exponential ones of random and WS type
networks.  One natural candidate is a power law, $n(k) \propto
k^{-\gamma}$.  It is a natural candidate because it has a very
long history in a wide range of subjects outside physics, as well
as playing a central role in several areas of physics.

For instance, Lotka \cite{Lot26} in 1926 claimed citations of
scientific papers followed a power law.  In this case the papers
are vertices and citations from one paper to another form edges.
This is a good example of where the internet communication and
desktop computer power has revolutionalised work on networks.
Redner \cite{Re} studied this network in 1998 using the 1991-1997
citations of 1991 papers listed in the ISI (Institute for
Scientific Information) database for his largest example.  This
gave him a directed network of about $N=7.8 \times 10^5$ vertices
with average degree around $K=8.6$. A much longer tail than a
simple exponential of this size and far more hubs are present. A
$\log(n(k))$ vs.\ $\log(k)$ plot reveals a linear tail for the
data with a slope of about $\gamma=3$.\footnote{As I have defined
small world graphs in terms of clustering and distance measures
relative to a similar sized random graph, there is no clear
statement one can make about the degree distributions of small
world networks.}.

Networks with power law degree distributions for large degree are
known as \emph{scale free networks} because ratios such as
$n(2k)/n(k)$ are constant, whatever the degree scale $k$ is chosen
to make the comparison\footnote{Always provided that we are
looking at vertices with large enough degree where $n(k)$ is a
power law.  The degree $k_s$ where power law behaviour starts does
in fact provide one scale for the problem.}.

\begin{figure}[htb]
\begin{center}
  \scalebox{0.8}{\includegraphics{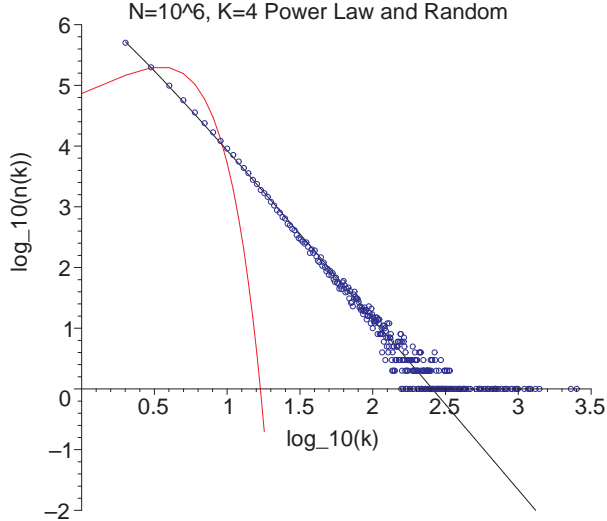}}
\end{center}
\caption{Log-log plot of degree distributions for a network with
$N=10^6$ vertices and average degree $K=4$. Data from a model of a
scale free network (blue circles) compared against a power law
with $n \propto (k+0.42)^{-2.82}$ (black straight line) (both with
minimum degree $k_0=2$).  Note how the power law crosses
$n(k_c)=1$ at about $k_c \approx 257$ ($\log_{10}(k_c) \approx
2.41$) and this is roughly where the data for the real network is
no longer continuous. Even if we used tricks to extract data for
higher $k$ the highest data point is likely to be around $k_1 \sim
4800$, though here it is at 2519, giving us only two and a half
decades to make the linear fit. Note that having a minimum degree
of $2$ rather than $1$ shifts the power law to the right compared
to the example with minimum degree $1$ considered in the text.
Also shown is the binomial distribution of the a random network of
the same number of vertices and edges (red curve) whose largest
hub is at $k \approx 17$. }
 \label{fspln1e6k4}
\end{figure}

To see that such a distribution has hubs, vertices with much
higher numbers of edges than in a random graph or in a WS model,
consider the following example.  Consider a network with $N=10^6$
vertices and an average degree of $K=4$ with vertices of all
degrees present whose degree distribution is of the form
$(k+k_s)^{-3.0}$ ($k_s \approx 2.91$) and so for large degrees its
just a simple inverse cubic.\footnote{If we specify $N$,$K$,
$\gamma$ and $k\geq k_0=1$ then we need one last free parameter,
say $k_s$, to allow a fit.  In this case we must have $k_s \approx
2.91$. In the following we analyse this distribution and its
various sums over integers by noting its simple relation to the
Hurwitz-Riemann zeta function whose properties are well known.} It
is easy to show that for every value between $1$ and about $k_c
\approx 284$ there is at least one vertex of that degree ($n(k) >
n(k_c) = 1$)
--- a continuous spectrum of degree. Once the likely number of
vertices with a given degree $k$ falls below one, $k>k_c$, then we
find that in a real network, where vertices have integer numbers
of connections\footnote{The formulae for degree distributions
$n(k)$ imply that some statistical average over many realisations
of networks is taken.}, the $n(k)=0$ for many $k
>k_c$ and we are at the end of the `continuous' part of the degree
distribution. However, unlike the random graph, the probability of
finding vertices with degree bigger than $k_c$ is not falling off
very fast and there is a a non-trivial chance of finding the odd
vertex with a large number of edges attached.  In this example we
expect to find about 143 of these large vertices with degrees
between about $k_c \approx 284$ and $k_1 \approx 1447$ where we
can show that there is unlikely to be a vertex with degree larger
than $k_1 \approx 1447$ ($\sum_{k=k_1}^\infty n(k) = 1$). Plots
for a similar distribution from a real model are shown in figure
\fref{fspln1e6k4}.
These numbers, $k_c,k_1$, are much larger than those for a random
network of same size where both are approximately the same and are
around 17. In fact for a power law distribution these numbers are
scaling as fractional powers\footnote{With the end of the
continuum at $n(k_c)=1/N$ and the largest degree around $k_1$
where $\sum_{k=k_1}^{N-1}n(k) = 1$, we find that $k_c \propto
N^{1/\gamma}$, $k_1 \propto N^{1/(\gamma-1)}$.} of $N$ while the
random graph cutoffs are roughly equal and scale roughly as
$\ln(N)$.

If the citation networks of Lotka and Redner were the only reports
of a power law distribution in the literature, then perhaps we'd
merely dismiss it as a special and rare case. However Newman
\cite{New01c,New01d} and others have reported similar patterns are
reported in scientific collaboration in different academic fields,
the modern version of the Erd\H{o}s number discussed above.
Albert, Jeong and Barab\'asi showed that \cite{AJB99,BKMRRSTW00}
that the network of links (the edges) between web pages (the
vertices) seems to be a scale free network. Three Faloutsos
brothers showed that the computers and routers that form the
internet itself form a scale-free network \cite{FFF} ($\gamma \sim
2.2$). Patterns in long distance phone calls \cite{APR99,ACL00}
($\gamma \sim 2.1$), the network of film actors used for the Kevin
Bacon game ($\gamma \sim 2.3$) \cite{BA}, the relationships of
words in English texts \cite{CS} ($\gamma \sim 2.7$) and many
other examples have been given in the recent literature. They all
have distance scales comparable to a random graph but much larger
clustering coefficients, i.e.\ they are also small world networks.
Indeed while not all networks are scale free, many in the
literature are reported to be just that.

A power law distribution is something that should again make all
physicists sit up.  At or near a second order phase transition,
many quantities follow a power law.  For instance the
magnetisation of a real magnet near the critical point is
proportional to $(T_c-T)^\beta$ where $T$ is the temperature,
$T_c$ the critical temperature.  Physically the scale invariance
reflects the fact that near a critical point the relevant
correlation length becomes infinite, there is no relevant scale
left. Further, this means the small scale details of the material
become irrelevant so one has universality, many materials with
different short scale interactions have the same behaviour near
the critical point. This insight led to powerful new ways of
solving problems near critical points. Subsequently, whenever a
power law is seen one hopes that this reflects the emergence of a
similar universality and so the hope is that if a simple model can
capture the essential physics, it will be guaranteed to give the
correct results for the real world examples, just as simple Ising
models are good models for the critical behaviour of real
materials.

Power laws are also central to more recent ideas.  For instance
the length of a coast line $l$ measured with a ruler of size $k$
is also a power law, $l \propto k^{-\gamma}$. In this case it
reflects the fact that the coast line is not a simple line of
dimension one but a \emph{fractal}  with a fractal dimension
$\gamma$.  In the 80's and 90's, power laws were at the centre of
interest in various complex systems. The distribution of the
number of earthquakes $n$ of a certain size (equivalent to $k$) is
a power law over several decades (Gutenberg-Richter law) even
though describing the properties of the earth's plates, their
response to forces, their interactions with the core, appears to
be far too complicated to give such a simple result.  A clue comes
from the existence of several models where the microscopic rules
are very simple, even if the equilibrium solution may be hard to
find analytically. Numerical or even real experiments can be
performed to show they show critical behaviour with power laws.
For instance in the sand pile models where `sand' grains are
dropped regularly onto a pile, with a rule that the sand is stable
unless the height difference between its neighbours exceeds some
critical point.  This produces a number of avalanches ($n$) which
is a power law of the size of the avalanches ($k$). One idea seen
in some models is that many systems may actually prefer to lie on
a critical point, one can alter the initial conditions or the
microscopic rules but they always give a power law if one waits
for some time. This is called SOC --- \emph{self organised
criticality} \cite{Jensen,Tur}.

Intriguingly, many power laws occur in areas outside physics, and
several famous ones predate physicists interests' in networks, SOC
or critical phenomena by a considerable time.  We have already
noted Lotka's interest in citations \cite{Lot26}.  He observed
that the number of authors contributing $n$ publications to a
bibliography is often an inverse square law, but did not look at
this in the context of a network (papers as vertices, citations as
links).

Many power laws in social science are said to be examples of
Zipf's law.  George Kingsley Zipf \cite{Zipf} noted that for many
quantities if you rank them in order of size, largest given rank
one, second largest rank two, etc.\ and plot their size against
their rank, then we get a power law.  One example he gave was of
the frequency of English words and he suggested the frequency of a
word was approximately inversely proportional to the rank. The
universality of this idea forms the basis of many data compression
algorithms. His second famous example is the sizes of cities and
again suggested that city size was inversely proportional to
rank.\tnote{It holds, with slightly different exponents in many
regions \cite{ZM97}.}
 \begin{figure}[htb]
\begin{center}
  \scalebox{0.4}{\includegraphics{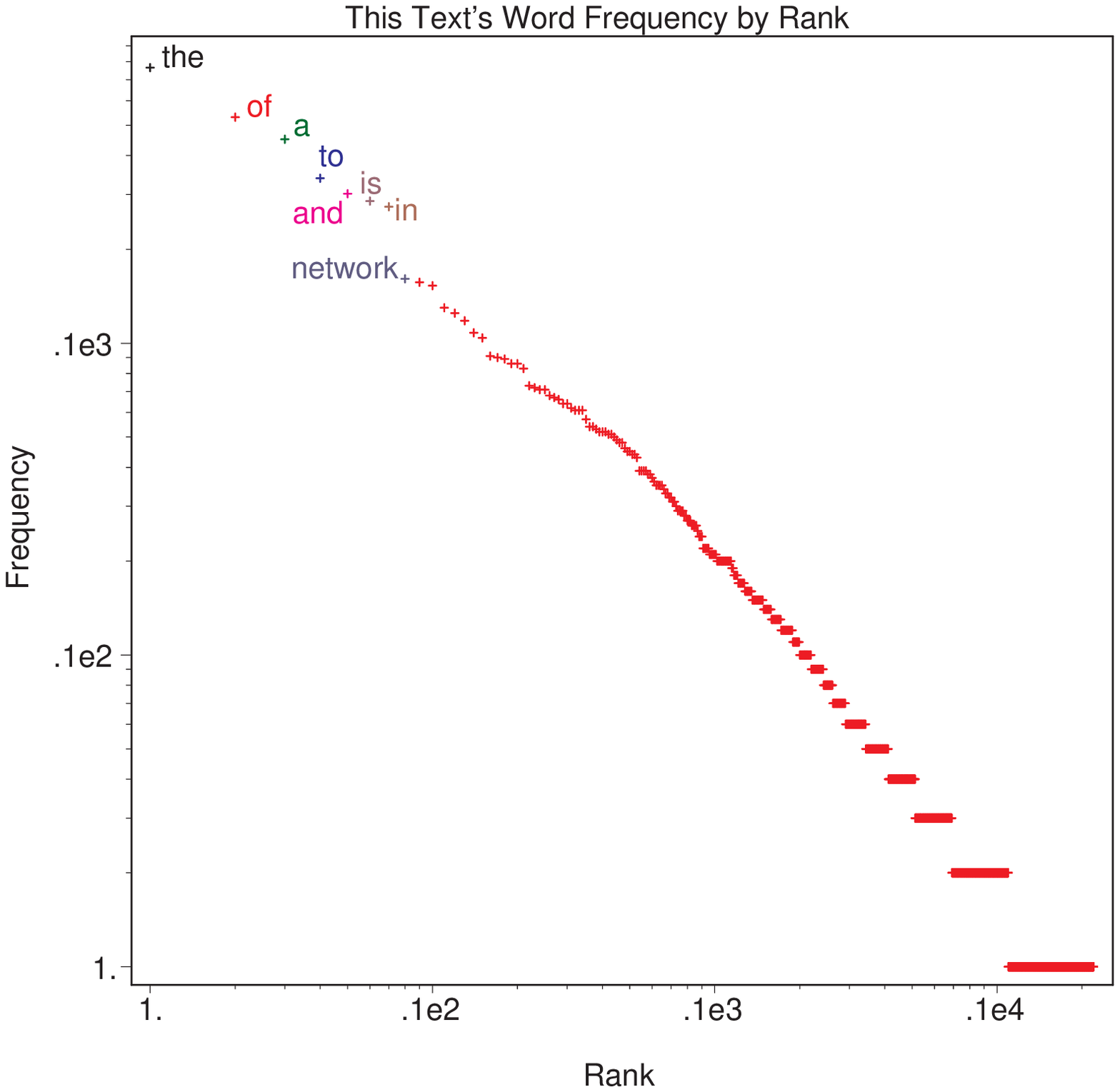}}
  \hspace*{1cm}
  \scalebox{0.4}{\includegraphics{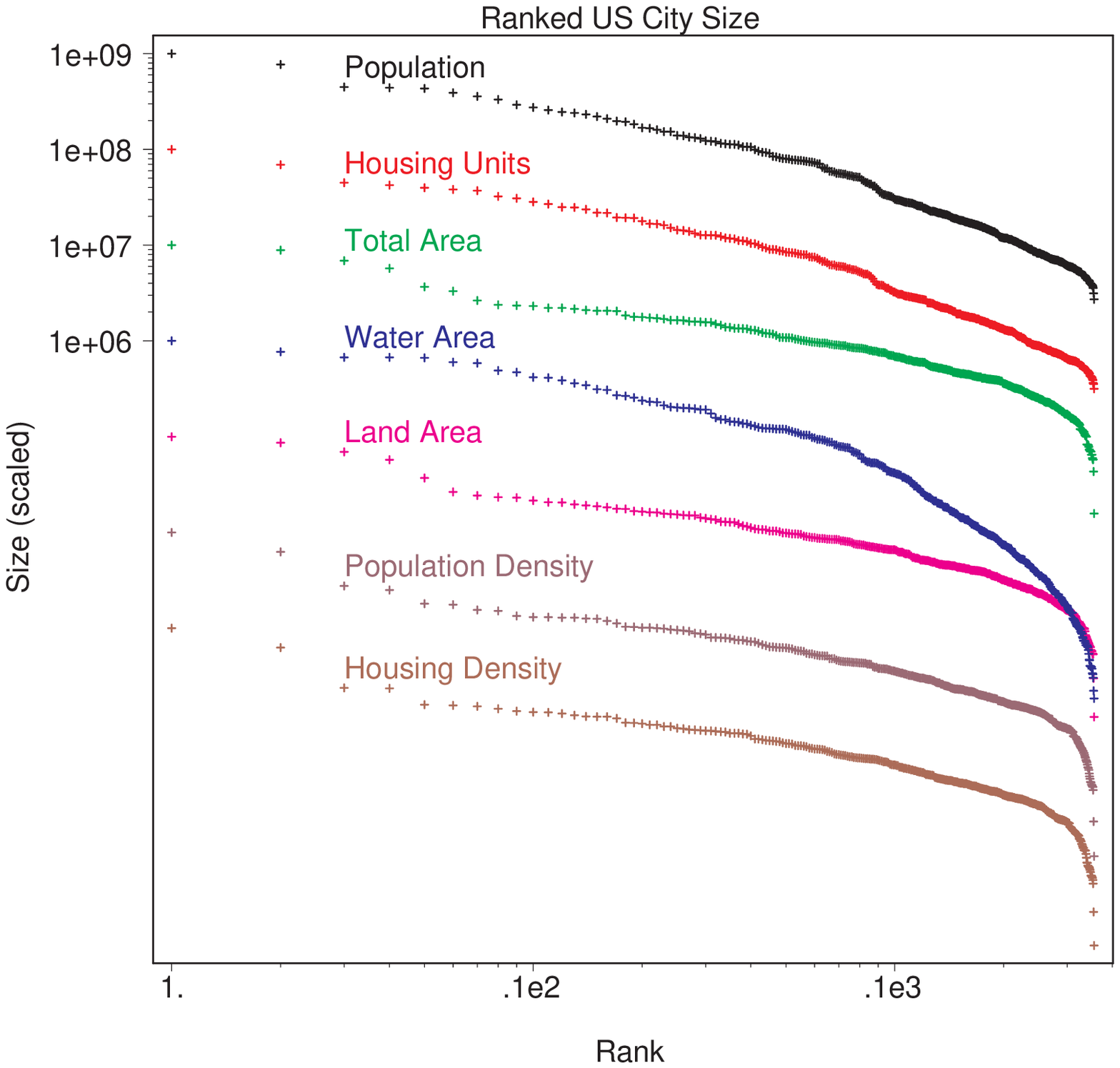}}
\end{center}
\caption{Two classic Zipf plots.  On the left a log-log plot of
word frequency against rank for this text ---  ``the'' is the most
common word, ``network'' is eighth. On the right, a log-log plot
of the size of US cities against rank, all data scaled relative to
the largest and by a power of ten relative to the next curve (for
visualisation purposes). The data is for ``Metropolitan Areas'' in
the USA in 2000 and taken from the US Census Bureau web site
\cite{USCB}. Size is given seven different ways in terms of
population, housing units, total area, water area, land area,
population density, and house unit density. Most show some
evidence of a simple power law.}
 \label{fuscitysize}
\end{figure}
We can turn our degree distributions into the language of Zipf's
law as the rank $r$ of a vertex of a network in terms of its
degree $k$, its `size', is given by $r=\sum_k^\infty n(k)$. Zipf's
law then consists of plotting $k$ against $r$. One can quickly see
that $k \propto r^{-\nu}$ with $\nu= 1/(\gamma-1)$ for a power law
distribution $n(k) \propto k^{-\gamma}$.

The characteristic of all these power laws can be summarised in
terms of an older `law', Pareto's law.  Pareto was an economist
working in the late 1800's who noted that 80\% of the land in
Italy was owned by 20\% of the population.  Such an 80:20 rule
applies whenever we have a distribution with a significant tail
such as a power law distribution.  For instance only a few English
words are needed to write most of a book, Lotka's study means only
40\% of authors have more than one publication in a bibliography.

Finally, one of nicest examples of scaling laws are those relating
to metabolic function observed in biological systems.  For
instance the lifespan of an organism scales as its mass to the
power one quarter and this relationship holds over twenty one
orders of magnitude, from microbes to whales.  It is particularly
relevant as the very ideas of scaling used so successfully in
physics to explain critical phenomena have recently been used by
West, Brown and Enquist \cite{WBE97,WBE99} to provide a simple
explanation for these previously mysterious biological scaling
laws.

So is it possible that when we see scale-free networks, networks
with a power law degree distribution, it reflects some simple
fundamental law that could be described by the network equivalent
of an Ising model? Are most of the details in network creation
irrelevant and we need only focus on some crucial ingredient,
something which means many networks will always eventually
organise themselves in a special way and the scale free degree
distribution is flagging this?  Are networks fractal in some
sense?  Not all the power laws above, such as the city data, have
an obvious network basis but perhaps we would gain great insight
if we could see their power laws as a result of some scale-free
network structure which we have missed to date. Many of these
power laws appear in areas well outside traditional physics,
social networks with their six degrees of separation and so forth.
Can the language of networks provide us with the first clues of
some rigourous theory of some aspects of human social behaviour
--- a realisation of Asimov's psychohistory \cite{Asi}?  Or is
this destined to remain science fiction?

\subsubsection*{A word of caution}

\begin{figure}[htb]
\begin{center}
  \scalebox{0.6}{\includegraphics{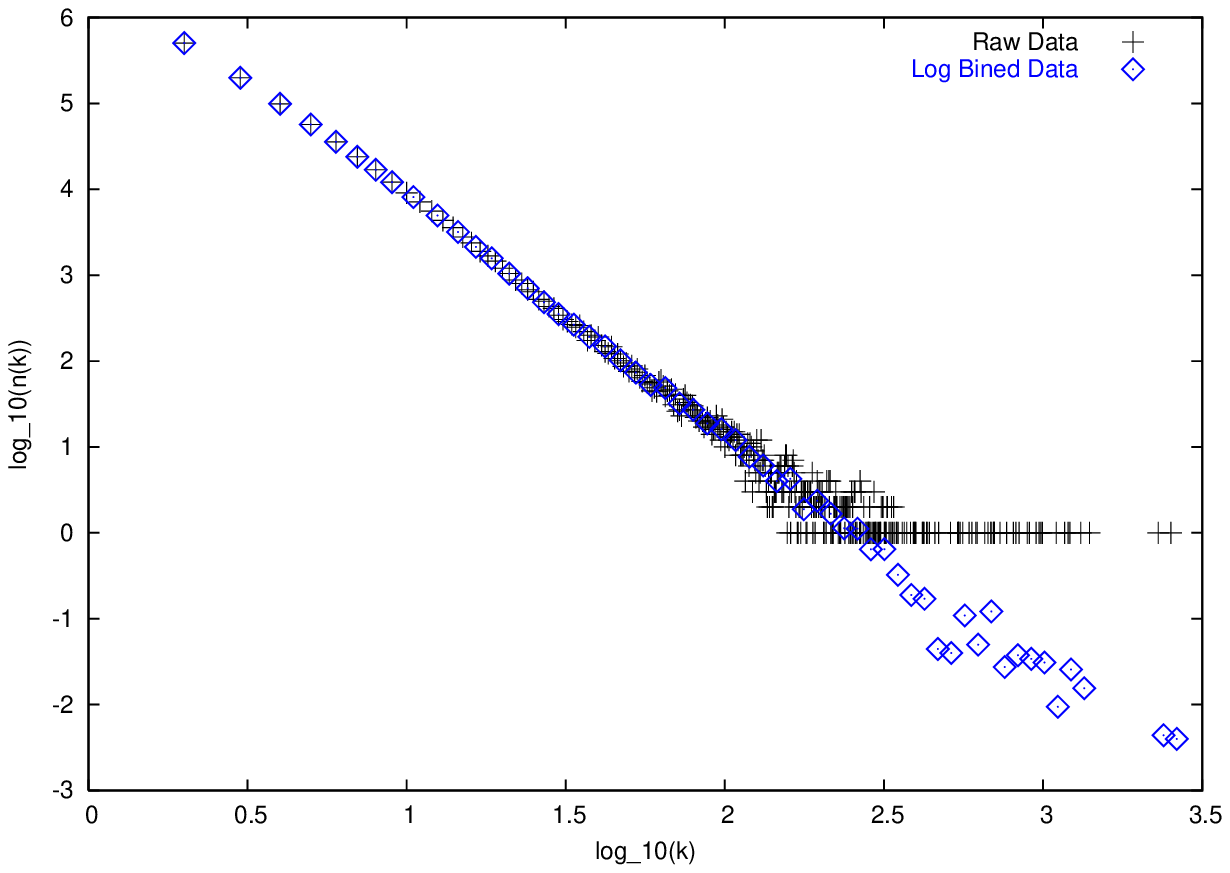}}
  \scalebox{0.6}{\includegraphics{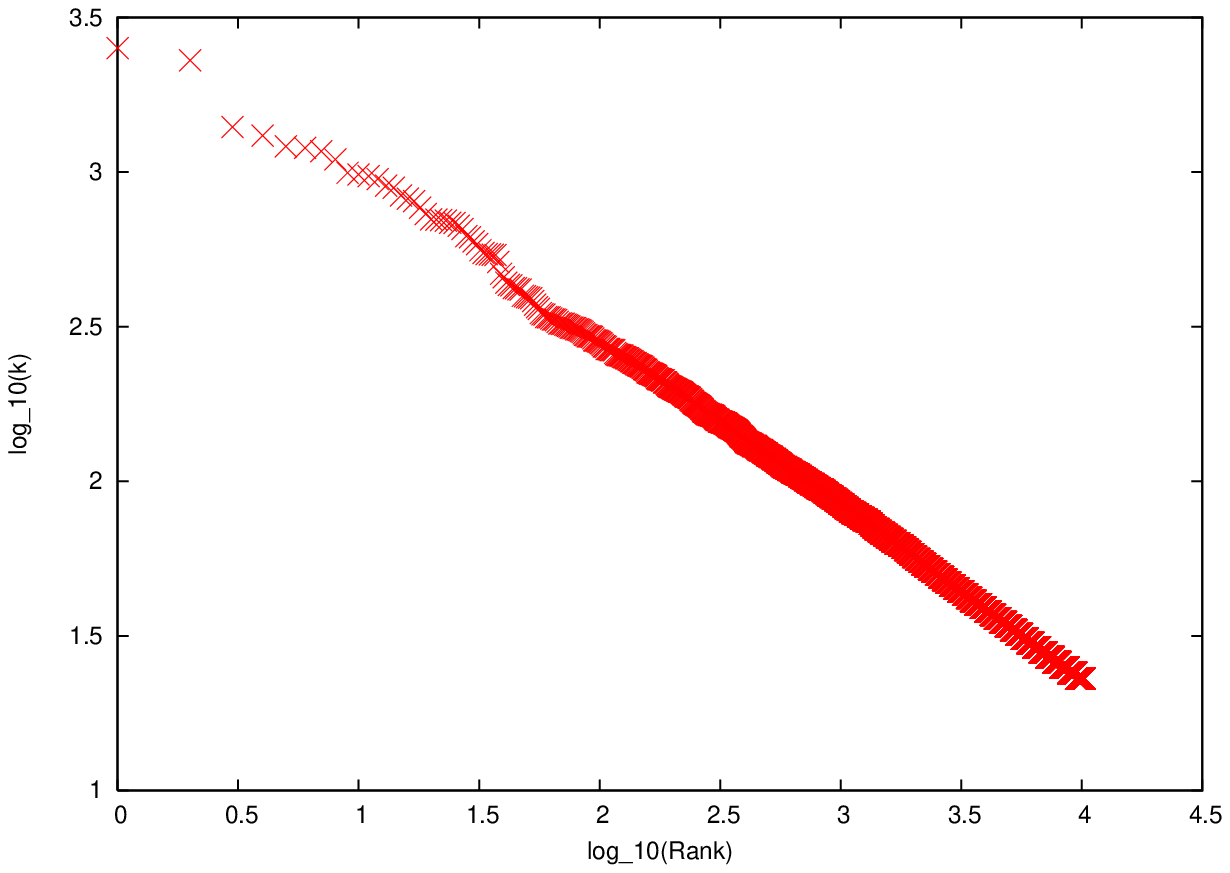}}
\end{center}
\caption{On the left a log-log plot of the degree distribution
from a theoretical model of a scale free network with $N=10^6$
vertices, average degree $K=4$, minimum degree $2$.  The best fit
gives a power of $2.82$. The black crosses are the raw data and
the blue diamonds are the result of logarithmic bining. Note how
the bining leaves the form unchanged from around $\log_{10}(k) =
1$ where it first makes a difference and has a consistent shape
until $\log_{10}(k) = 3$ though the quality of any fit will be
poorer at these high $k$. On the other hand the raw data has large
fluctuations from around $\log_{10}(k) = 2$ so the bining is
gaining us nearly a decade of data to fit any theoretical curve
to.  As an alternative, the log size (degree) vs log rank plot \`a
la Zipf also shows a much cleaner result for the biggest $10^4$
vertices with slope around $-0.55$.}
 \label{fbining}
\end{figure}

These ideas about power laws, these questions, these speculative
links certainly provide one explanation for the excitement in the
study of networks in recent years. However, a word of caution.
Modern computers and the internet have allowed researchers to
create and analyse large databases across a wide range of topics.
It may seem that a data set of $10^6$ vertices is large (and I can
think of only three examples which are much bigger). However the
degree distribution is much shorter than $10^6$. Suppose our
system gives a network of $N=10^6$ vertices, average degree $K=4$
and a minimum connectivity of $k_0=2$ and a $(k+k_s)^{-\gamma}$
form for the degree distribution with power $\gamma=3$ and scale
$k_s = 0.87$ so its effectively a pure power law for almost all
$k$ . One typical example of such a network will have a non-zero
degree distribution $n(k)$ only for $k$ below $k_c \sim 226$
(where $n(k_c) = 1$). Above this value there are often no vertices
with that particular degree $k$, $n(k)=0$, with only the
occasional $n(k>k_c) =1$.  So plotting this directly on a log-log
plot we have to ignore many points where $n(k)=0$ while the few
remaining ones are all the same value, one.  For similar reasons
the region just below $k_c$ also shows large fractional
fluctuations. The data from a computer model of a similar
scale-free network in figure \fref{fbining} shows this clearly.
Thus even in the best case where there is a power law for all
values of $k$ (scale free networks need be power law only for
large $k$) we have only two decades of linear behaviour in our
log-log plot. Anyhow, in most cases the power law form starts at
some degree higher than one.

There are various tricks which improve the situation a little. One
is to bin the data in the large degree $k \gtrsim k_c$ region as
in figure \fref{fbining}. That is if we count $\eta$ vertices of
degree between $k_\mathrm{low}$ to $k_\mathrm{high}$, one `bin',
we assign a single data point of value $n(k_b) = \eta
/(k_\mathrm{low}-k_\mathrm{high})$ at\footnote{The optimal
position chosen depends on the nature of the curve, $k_b =
\sqrt{k_\mathrm{low}k_\mathrm{high}}$ is another choice but in
practice the two are essentially the same for
$k_\mathrm{low}/k_\mathrm{high} \ll 1$.} $k_b =
(k_\mathrm{low}+k_\mathrm{high})/2$. If we choose these bins so
that $k_\mathrm{high}/k_{low}$ is a constant, we get a series of
equally spaced points on a log-log plot.  By averaging over a
range, we have fewer data points but they no are longer limited by
the discrete values $1$ and $0$ and they allow us to fit in this
region. However, with the largest degree around $k_1 \sim 2400$,
$\log_{10}(2400) = 3.4$, we will not add more than a decade to our
linear region.
As an alternative, one can try to plot a Zipf-like size vs.\ rank
distribution, see figure \fref{fbining}.  However, both bining and
ranking introduce correlations between data points so the
statistical analysis is not so simple.

It is not surprising then that many authors do not put an error on
the power law fitted to the data, nor can one always exclude other
types of fit to the data, stretched exponentials $k^{-\gamma} \exp
\{ -(k/k_s)^\beta \}$ for instance. It would be nice to get more
data but it is only with modern computers that networks of a
million vertices can be easily found and, even then, I know of
only two studies of the web and one of phone calls where the
number of vertices is larger (though still no more than $10^8$).
Many data sets are smaller and are not going to get significantly
bigger. The history of power laws in other areas of modern physics
tells us that they are so intriguing that we sometimes rush to see
power laws in everything and this is premature in some cases.
Another consequence is that we are not going to see power laws
with large coefficients \cite{DM01}.  If we require at least two
decades of linear data and assume a simple power law for the whole
range of $k \geq 1$ then we require that $k_c = 100$, where
$n(k_c)=1$, and we need $N \sim 100^\gamma$ vertices.  Thus with
$N = 10^6$ we should be able to see scale free behaviour in
networks if the power $\gamma$ is less than 3. So while the
ranking and logarithmic binning can help improve the data, the
power law section of accurate data is rarely going to be even this
long we would should expect three to be an upper practical limit
on powers found in real data.\tnote{See \cite{DM01} review, page
27.} The worry is that with only short ranges of linear-like data,
and two decades is not so long, many other forms can fit the data
just as well, such as stretched exponentials. These additional
forms can not be excluded. If we are interested in the form the
network would have as $N \rightarrow \infty$ then we must be very
careful, as finite size effects may be difficult to detect with
the data. On the other hand if we are just characterising our
network, and not ascribing some deeper meaning to an exact power
law relation then we need not be so worried. Just by looking by
eye at many of the data sets, one can be convinced that the
$\log(n(k))$ vs.\ $\log(k)$ plots do have convincing linear tails
and the identification of scale-free networks in a wide range of
networks in the real world is not unreasonable.

\subsection*{So What?}

So far we have discussed several key aspects of a network which
allow us to distinguish different types of network.  While
classification is important, let us now try to see if there are
further useful questions we can answer about networks.

\subsubsection*{Optimisation}

There are several types of question asked about traditionally
asked about networks. There are the `optimal route' questions such
as the travelling salesman problem mentioned at the start of the
article.  Perhaps the oldest example is the one that is seen as
the start of graph theory in mathematics.  In 1736, the great
Swiss mathematician Euler proved that it was impossible to walk
round the city of K\"onigsberg (modern day Kalinigrad) crossing
each of its seven bridges across the River Pregel once and only
once, a preoccupation of some of its citizens at the time. This is
equivalent to asking if you could walk along every edge of a
particular network once and only once. Critical path analysis is
another well established problem. If we use a graph to represent
the different tasks in a problem, each depending in some way on
earlier parts of the problem to be complete as indicated by
directed links, can we find the bottle necks in the process so we
can focus our resources on these and complete the job in the
shortest possible time. Problems such as these have a long
tradition and finding algorithms which give one good answer in a
reasonable time, rather than finding the best answer at any cost,
preoccupy many computer scientists.  The application of such
optimisation problems to a wide range of problems predate the
current interest in networks. For instance Davis \cite{Davis} used
network methods to suggest that the pivotal role of the island of
Delos in ancient Aegean culture, when it is a rather small and
insignificant island, was due to its critical position in the sea
borne communication networks of the era.

However, let us look at some of the questions that the recent
interest in networks has provoked or revived.

\subsubsection*{Where do networks come from?}

One of the first questions that comes to mind is why do different
types of network appear in different contexts?  How are different
types of network formed?  This may be related to the origin of
some of the power laws found in  a wide range of human experience,
city sizes by rank, river size against river basin areas, the
Gutenberg-Richter law of earthquake size-frequency, etc.\   Is
there a simple guiding principle behind the patterns in systems
with such complex interactions and can this be related to some
type of network? Here there is real hope. The story of the
elucidation of the simple physical and biological principles which
can explain the power laws seen in biology \cite{WBE97,WBE99}
offers an exemplary model, at least the way most physicists view
problems.

For the lattice we have regular solids with short range potentials
and a minimum energy principle to guide us and these do not
interest us here.  The random graph and WS models were presented
in terms of an algorithm for their creation. Thus we can see how
small world networks might be created. However, are most networks
going to be formed by purely random interactions (rewirings)?
While there may be an element of chance about which web pages a
web page author reads and therefore which links that author is
likely to add to their pages, there is also a large amount of
informed decision making going on.  The author reads web pages
related to their interests not just any old page and they link
only to such pages in general. Thus the Erd\H{o}s and Reyn\'i,
random graph model and the Watts and Strogatz algorithm for small
worlds while useful for theoretical analysis are perhaps not good
models of the processes that lead to networks in the real world.

\begin{figure}[htb]
\begin{center}
 \scalebox{0.5}{\includegraphics{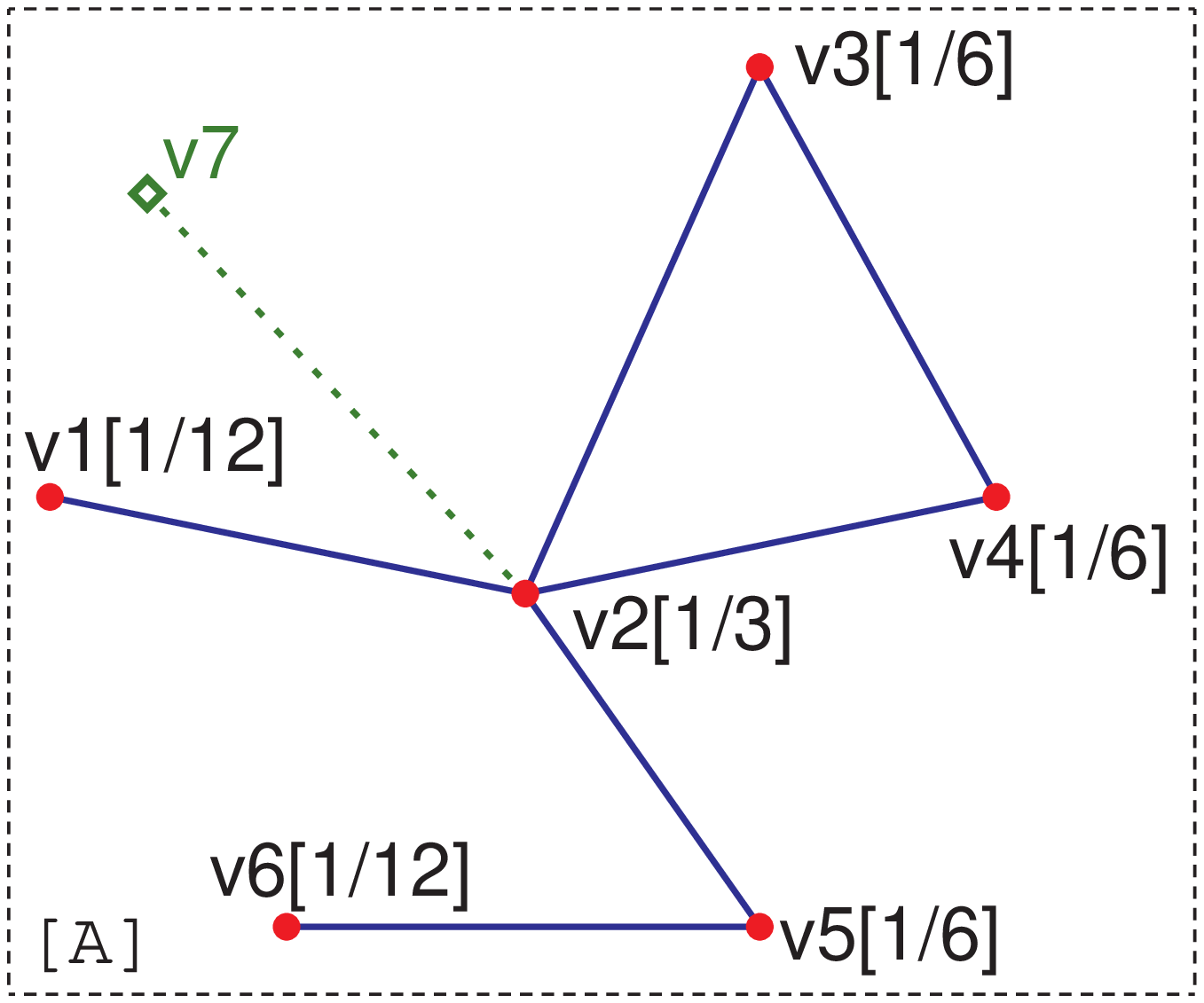}}
  \hspace*{0.5cm}
  \scalebox{0.5}{\includegraphics{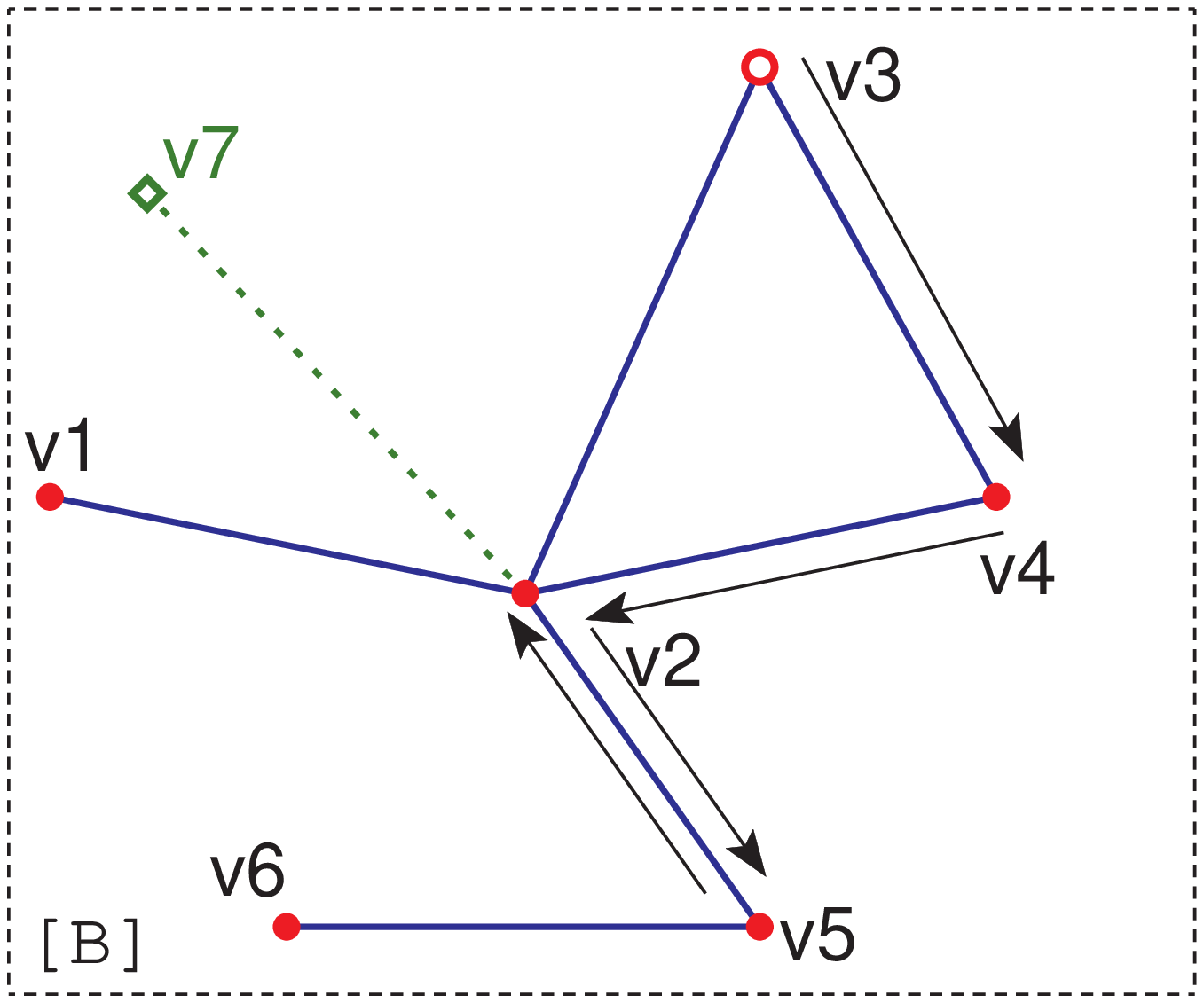}}
\end{center}
\caption{Two of the algorithms for creating scale-free networks.
In both cases, we add one new vertex v7 (green diamond) and want
to add one new edge between it and one of the vertices in the
existing network, v1 to v6.  The most likely edge to be chosen is
indicated with a dashed green line and is to the one with highest
degree, v2, in both cases. On the left, [A], we illustrate the
algorithm discussed by Barab\'asi and Albert in which we connect
our new edge to an existing vertex with probability proportional
to the degree of that vertex.  These probabilities are given in
square brackets.  On the right, [B], the walk algorithm is
illustrated.  We choose at a vertex from the existing network at
random, say v3, and then start a random walk from this point.  A
typical example is indicated by the arrows.  We stop after a
certain number of steps and connect the new vertex to the vertex
at the end of our walk, v2 in this case.}
 \label{fsf}
\end{figure}

One significant recent contribution has been to provide a model
for the creation of scale-free networks.  As we have seen, these
have been of interest for many years, though usually interest
focused on the power law and not any underlying network.  One can
find algorithms for creating scale-free networks e.g.\
\cite{ACL00},\tnote{Cited p6 of \cite{BS02}.} which involve the
input of the power law form but this is not going to help us study
their origin in the real world.  A major insight came from
Barab\'asi, Albert and Jeong shortly after the Watts and Stogatz
work on small worlds.  Their study of the world wide web
\cite{AJB99} showed a network with lots of hubs, incompatible with
random graphs. Seeing that the web was growing at a considerable
rate, they suggested a model that had two key ingredients, growth
and what they called `preferential attachment'.  Imagine adding
one new vertex at each turn to an existing network.  To this
vertex we attach $K/2$ new edges with one end connected to the new
vertex, the other end connected to an existing vertex chosen with
probability $\Pi$ from all the vertices in the existing network.
If $\Pi = 1/N$ so that all existing vertices are treated equally,
we end up with an `exponential' network\footnote{Not a random
graph of Erd\H{o}s-Reyn\'i type but a network with a simple
exponential fall off for the degree distribution.} and there will
be no hubs, no vertices with far more edges attached than the
majority. To reach a scale free model Barab\'asi and Albert
\cite{BA} suggested that these new edges attach preferentially to
vertices with large degree $k$. A ``rich get richer'' algorithm,
echoing Pareto's law. The simple form they took was $\Pi(k)
\propto k$ where the normalisation is simple to calculate, as
illustrated in figure \fref{fsf}A.
What is remarkable is that this model gives a power law degree
distribution with $\gamma=3$. In fact this model is a simpler
version of models suggested by Simon as early as 1955
\cite{Simon55,Simon57,BE} and there are several simple variations
of this model, such as $\Pi(k) \propto k + c$ with $c$ constant,
which can give scale-free networks of any power above\footnote{A
pure power law with a power equal to two has formally an infinite
average degree, informally too many edges. This is why models
can't reach powers of two and below.} $2$. Moreover, many simple
models can be solved exactly in a mean field approximation (see
for example \cite{BAJ,KRR,DM01,AB}).

Unfortunately what is also remarkable is that in this type of
model \emph{only} probabilities $\Pi$ linear in degree give power
law networks.  Anything else gives stretched exponential, such as
$n(k) \sim k^{-\gamma} \exp \{-(k/k_s)^{(1-\gamma)}\}$, or other
forms as seen in explicit solutions \cite{KRL}. It seems unlikely
that we choose which web pages to link to with probability
\emph{precisely} proportional to the number of web pages on a
site. Still the Barab\'asi and Albert model has highlighted
several crucial aspects about scale-free networks and power laws.
In particular it emphasises that the power laws come from networks
where there are hubs, that is vertices with far far more edges
attached than the vast majority and far bigger than found in
simple random or WS type small world networks. The idea then is
that the processes leading to the formation of a scale free
network \emph{require} some type of `preferential attachment', a
preference for the most connected vertex, does make qualitative
sense. If I think it is worth linking to a web site, the chances
are several other people will think the same.  I tend to link more
to popular web sites. The more popular sites are those to which
many different people are linking to. It is perfectly reasonable
to think that we are more likely to link to popular web pages than
obscure ones, and that popular ones are the ones with the highest
number of links to them, the highest degree.

In fact the idea can be developed to produce more realistic models
for scale free networks \cite{Vaz00,Vaz02,Kl02,SK04}.  Suppose I'm
writing a new web page and want to add a link to good pages about
my favourite hobby. What I might do is use the web to find these
pages.  So I go to a search engine (a super hub surely?) which
points me to various pages. I try a few, perhaps follow links from
those to pages now distance two from the search engine vertex.
What I am doing is walking along the edges of the network.  It is
the structure of the network which guides my search and thus I
would expect that this guides the way I connect to it, the way the
network grows. Of course, what happens is I am most likely to end
up at sites that are linked to by many others. There are more ways
of getting to them than the sites with only one or two links from
the outside world. The very search I do is likely to lead me
preferentially to the most connected sites. Indeed, if we idealise
this, and suppose that we execute a random walk on the network,
assuming we walk for a distance $d \gtrsim L$ the mean distance, I
am likely to arrive at a vertex independent of my starting point
and the probability of me finishing at a vertex with degree $k$
will be proportional to $k$ as there are $k$ different ways of
arriving at this vertex\footnote{Strictly speaking this is using a
mean field approximation.}.  Thus if we were to attach a new
vertex to a vertex in an existing vertex where the latter is
chosen after a random walk on the lattice, we will quickly
generate a scale-free network. Such a `walk algorithm' was used to
generate the scale-free models in figures \fref{fspln1e6k4} and
\fref{fbining} and is illustrated in figure \fref{fsf}B.

Growth is also not a requirement for scale-free networks, one just
needs an on going dynamical rearrangement. One approach is to
rewire as in the original WS model but now we choose how to rewire
an edge in a more general way, e.g.\ choosing to reconnect
existing edges to new vertices of degree $k$ with probability $\Pi
\propto k$ \cite{EW02}.  More generally we can steal ideas from
statistical physics and define a Monte Carlo type algorithm in
which the network emerges as an equilibrium state. The idea would
be that we give the network an `energy' with the Hamiltonian
containing terms proportional to the number of vertices, the
number of edges and other suitable terms. We would try removing or
adding an edge or a vertex, and use the Monte Carlo algorithm to
accept or reject such an update. The coefficients of the terms in
the Hamiltonian play the roles of temperature and chemical
potentials and we could imagine fixing some of these terms, the
equivalent of working in different types of ensemble.  The
equilibrium solution gives the number of vertices with a given
connectivity rather than the number of states with a given energy
and this allows us to see which type of terms in the Hamiltonian
are required for different types of network.  In the real world,
such a Hamiltonian may arise because we have costs, each edge of
the internet might be a cable which costs money to use.

\subsubsection*{Desperately seeking ...}

In the quote from Guare's play, I left out a couple of sentences.
Of relevance here is a comment following the first part on the six
degrees of separation
\begin{quote}
``I find that A) tremendously comforting that we're so close and
B) like Chinese water torture that we're so close.  Because you
have to find the right six people to make the connection.''
\end{quote}
while the last part continues
\begin{quote}
``But to find the right six people. ''
\end{quote}
Guare is highlighting an important aspect of Milgram's experiment,
namely how we search the network to find the shortest paths. If
there were only six degrees of separation between people in
Nebraska and Massachusetts, how did they find this route? There
are infinitely many ways that a letter could have been passed
along.  Indeed the fact that only about 20\% of the letters sent
actually arrived could be in part be due to the fact that some
people did not find a good route. Further, perhaps there were
shorter routes but people were not able to find them. There is not
much point in having a network if we can't find our way about it.
Finding good routes will be an essential part of almost any
network. In anthropological terms, many cultures set up a network
of contacts, for example through marriage or gift giving. One use
of such a network is the right to use it to find a skill or
resource you need but don't have.  In terms of computing, one way
of increasing computer power is to distribute the computing and
storage facilities across many computers.  The peer-to-peer file
sharing networks can be run without central coordination but then
how do you find which computer has the file you want?  If one
sends out a message to all your neighbours in the network, and
they in turn send out to all their neighbours if they don't have
it, the network will be swamped by such requests
\cite{ALH02}.\tnote{p308.}

Consider for example Kleinberg \cite{Kl00} who considered a
two-dimensional lattice to which short cuts have been added with
probability proportional to $d^{-\alpha}$ where $d$ is the
Manhattan distance (network distance if there are no shortcuts).
He showed that if you know the Euclidean position of the target
and if the short cuts satisfied an inverse square law $\alpha=2$
then with a simple algorithm he specified which used only
knowledge available locally at each vertex\footnote{Take the edge
which decreases the Manhattan distance to the target by the
maximum amount.}, it was possible to find short paths from initial
to target vertex, and it would take about $(\ln(N))^2$ time
(equals the number of steps taken on the network).  For any other
situation, a different algorithm, different power law for the
short cuts, the time taken would be considerably longer.

Kleinberg's work is a good example of what can be achieved
analytically and it is relevant for many problems.  However
knowing that there is an underlying lattice and using the
Manhattan distances is in a sense using some global information
about the network.  In Milgram's experiment, this might have
helped to some degree, a letter from Nebraska might have been sent
first to a friend in Boston, as this is physically closer to New
York, who then forwarded to a friend in Massachusetts. However
once in the final geographical region, another way of finding the
target is needed as only rough geographical information was given.
Presumably the searches in this case were then made by using
intuitive ideas about the distance between different professions
or something similar. For instance if the target is a medical
doctor then I give the message to any friends who are medical
doctors or, failing that, I give it to any one who works in the
medical profession as while they are not as `close'
professionally, its natural to think of them as `closer' than
other friends who are car mechanics. In this case people are parts
of several overlapping networks, networks of Euclidean position,
networks of professional relationships, and so forth. Vertices
which are distant, because of links in one network, may be much
closer in another network.  Thus it is by being part of several
unrelated networks that enables this search to be successful in
relatively few steps.  Finding good algorithms under different
circumstances is one of the major activities in this area
\cite{WDN02}.  It has also prompted new attempts to gather
experimental information on such networks and searches for example
\cite{DMW03}.

\subsubsection*{Groups}

One area where there has been a great interest in networks, and
one which predates Watts and Stogatz paper, is in areas we might
broadly label as anthropology. As Milgram's experiment highlights,
data on human interactions is much harder to collect yet it often
has immediate relevance to us all.  One should not be surprised to
find that ideas from graph theory have been borrowed and developed
by researchers in this field for many years.

One question that is often asked is, what are the groupings or
cohesive blocks in this network?  If you ask people or
organisations to whom they are connected, to which groups they
belong, what alliances they have, the trouble is they may well
give you the acceptable answer for their society or business world
and not the true one. Indeed they may not even be aware of which
links are the most important. Gathering the information in
whatever way one can is often challenging in this field, but given
the data the idea would be to see if the network structure itself
gives us unbiased identifications of the real groupings. With this
information many other aspects of the way such system works might
become more obvious.

\begin{figure}[htb]
\begin{center}
  \scalebox{0.45}{\includegraphics{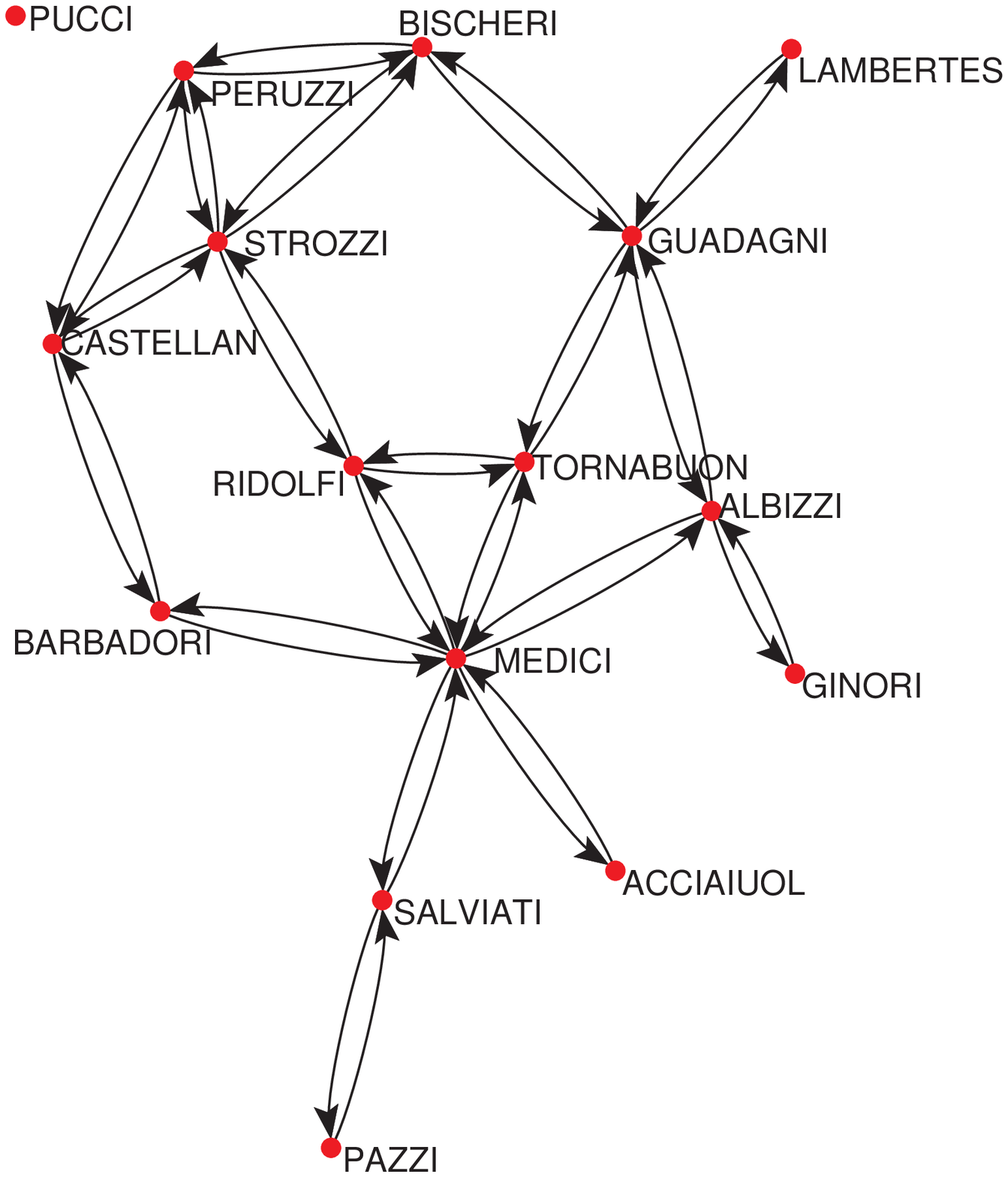}}
  \hspace*{-2cm}
  \scalebox{0.45}{\includegraphics{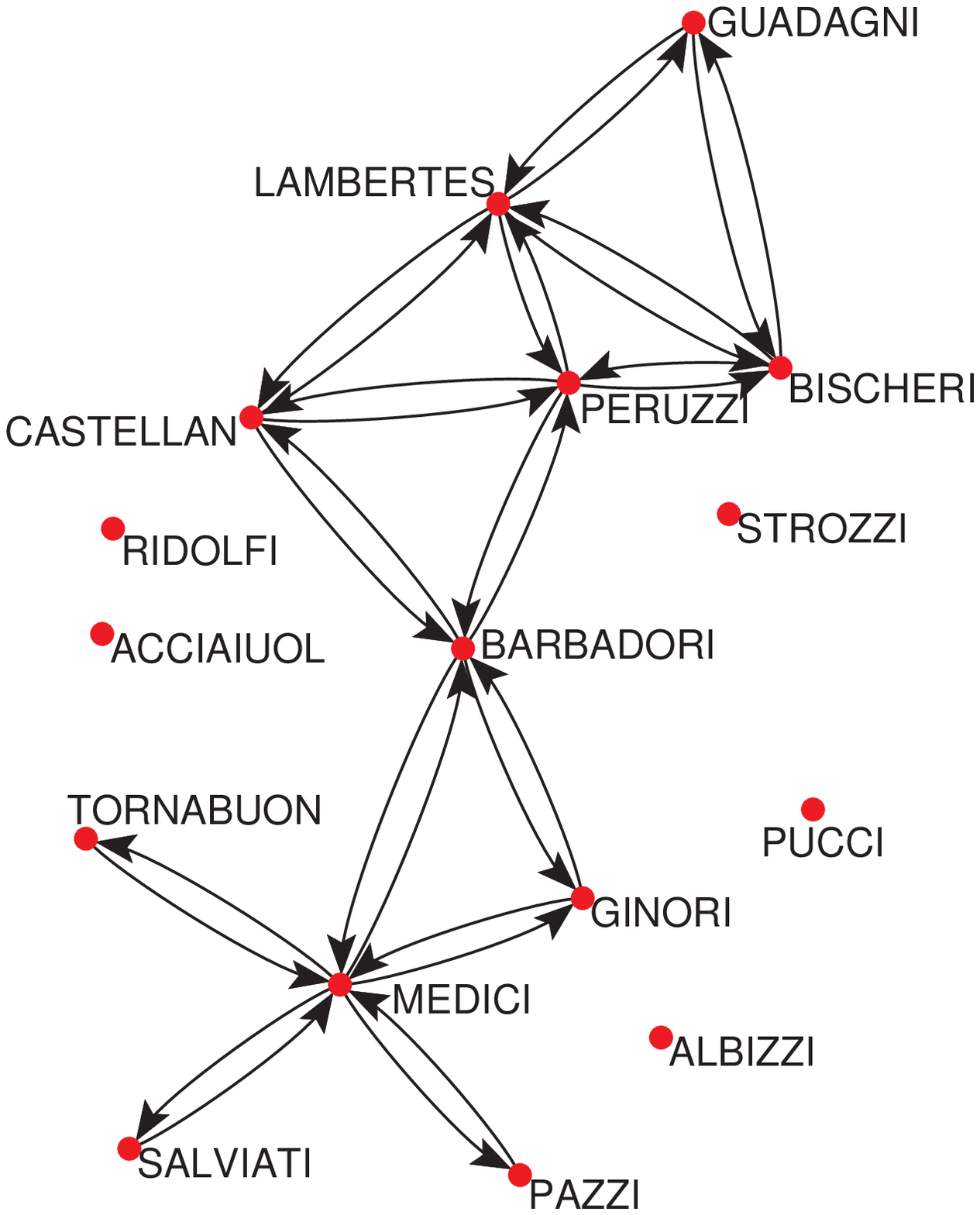}}
\end{center}
\caption{The networks between Florentine families around 1434,
marriage on the left, business on the right. Data based on the
work of Padgett and taken from examples provided by UCINET
\cite{UCINET}.}
 \label{fflorence}
\end{figure}

\begin{figure}[htb]
 \begin{center}
 \scalebox{0.6}{\includegraphics{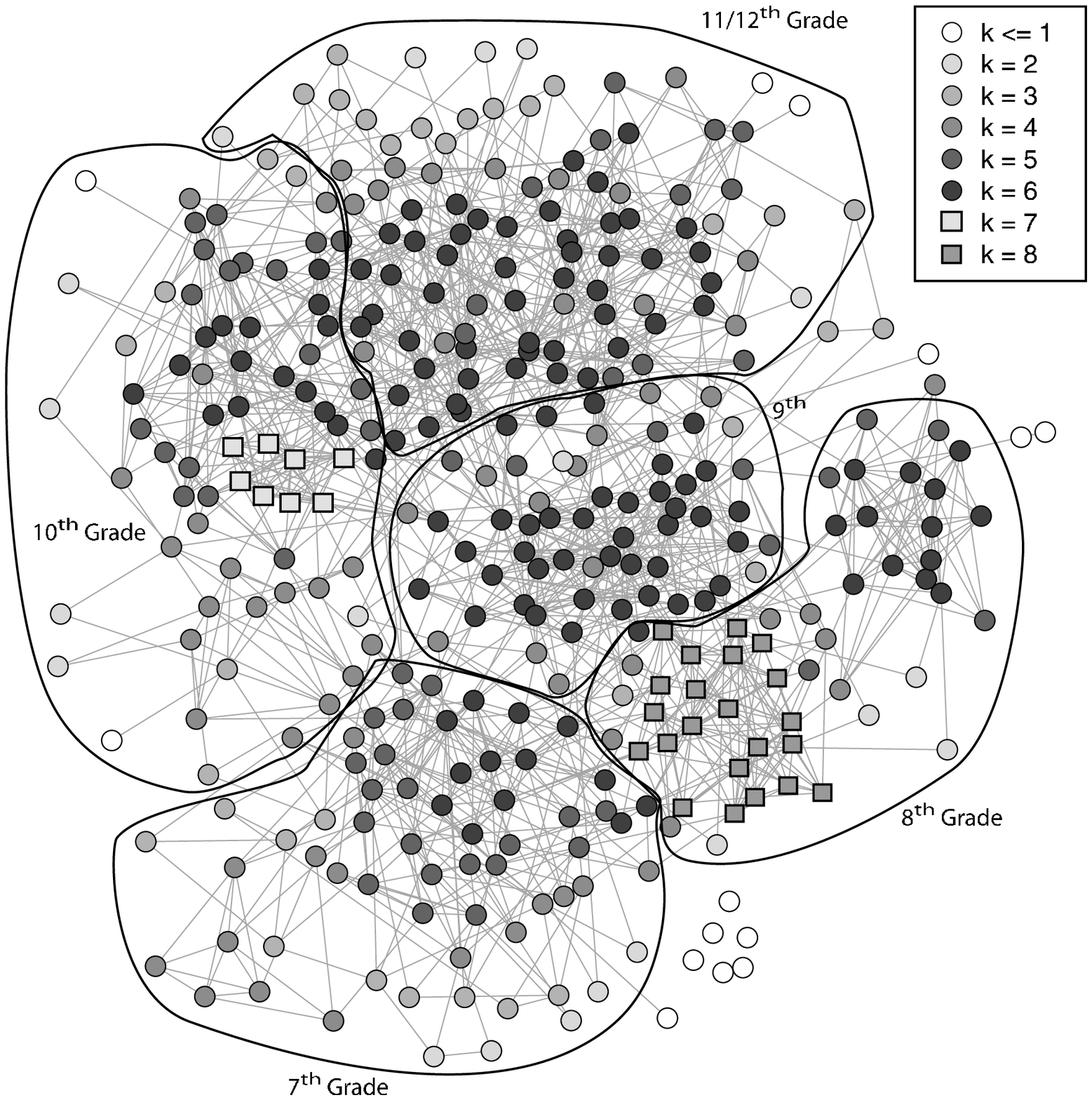}}
 \end{center}
 \caption{The network of social links in an American High school
analysed by Moody and White using network ideas.  Nodes are
students and their relationships are links. The vertices are
distinguished by their degree $k$ as given in the key. The
cohesive groups overlap in `k-ridges' with components centred on
organisation by grades. Their interpretation is as follows:
7th-graders- core/periphery; 8th- two cliques, one hyper-solidary,
the other marginalised; 9th- central transitional; 10th- hang out
on margins of seniors; 11th-12th- integrated, but more freedom to
marginalise.}
 \label{fhighschool}
\end{figure}

Take for example the work study of families in 14th century
Florence by Padgett and Ansell \cite{PA93}.  By expressing the
relationships between families as edges in a network, figure
\fref{fflorence}, a distinctive pattern emerges where Cosimo de
Medici was at the centre of the block under his control, whereas
his rivals for power had a much more diffuse structure.  The
suggestion is that it is the very structure of the network around
the Medici family which led to Cosimo taking over control of
Florence in 1434. So whether created  by design or as a reaction
to external pressures, if one had studied this network, one might
have been able to even predict the success of Cosimo.
Note though that these data sets are far smaller than those many
physicists discuss.  It is completely meaningless to try to label
this as a random graph, small world  or a scale free graph, etc.,
it is just too small.  Rather network based ideas are used to
identify small groups and cliques.  In this case the ideas of
block modelling \cite{WBB76} was used.   These ideas have been
extended, for example in Moody and White \cite{MW03}.  They study
the network of social links in an American High school using
measures of `cohesion' and `embeddedness' derived from graph based
concepts such as k-connectivity and applying Menger's theorem.
They test their results against `outcome variables'. Their groups
match the formal organisation of the school in terms of year
grades but reveal additional details, such as the tendency of
younger grades not to be fully assimilated into the central group,
and older pupils having more confidence to stay outside the
central group, see figure \ref{fhighschool}.

In a similar way, Page and Harary \cite{PH91} used networks to
express the different types of information known about the
different islands and people in Oceania.  These are often based
upon exchange of good, services,  and information.  Again one can
gain some understanding of how these cultures operated.

\subsubsection*{Dynamical Games on a Network}

Finding a network in the real world, establishing its nature, is
often only the precursor to the real question.  Often it is not
the network but what happens on top of the network that we are
really interested in.

For instance, it may be useful to know how computers, switches
etc.\ are connected to form the physical backbone of the internet,
but only once we know its structure can we ask how vulnerable is
it to random failure or malicious attacks \cite{AJB00}. As we have
noted the internet does appear to be scale-free and small-world
\cite{FFF}. It was designed by Baran \cite{Baran} to have a high
level of redundancy so it could withstand a large amount of random
damage, because of the original military applications. The many
short cuts present in a small world ensure that this is indeed a
feature of such networks.  If we were to pick a vertex at random
and remove it, most paths would lengthen a bit but basically there
should be no big problem. Certainly, every vertex would still have
a path to all other vertices, the network remains connected.  On
the other hand, if there was a deliberate attack which removed
vertices with the highest degree, the hubs, then a scale-free
network will quickly split into several disconnected pieces as so
many routes go through a few large hubs \cite{AJB00}.  A graph
with no hubs, a random or WS small world network, does not suffer
unduly under such targeted attacks and still has relatively short
paths.

A related topic is virus spreading and immunisation, be they
computer or human viruses.  It is essential for our understanding
that we use the right network of contacts when we simulate
transmission of viruses.  Interestingly, simple models of virus
transmission often show a transition from no infection to some
infection as immunisation levels are lowered (e.g.\ SIS models).
However similar models of virus spreading on scale-free networks
often show no such transition. Again, there are too many paths
linking one vertex to another \cite{PV,PV02,New02b} so that a
virus can always find ways of reaching uninfected and unprotected
individuals.  On the other hand, using the analogy of the
malicious attacks on a scale-free network, its clear that
immunising the hubs, the vertices with most connections, should
bring great dividends.  For those whose email is blighted by
viruses and spam have probably already realised that immunising
random vertices is pretty ineffectual.  It only takes a few
unprotected computers, a few sites prepared to host the traffic
for the infection to spread on a scale-free network. To stop this
you have to put the burden on the hubs.

This can be taken much further.  Just as with biological virus
spreading, there has been interest in all sorts of models of
individuals interacting on a small scale to produce large scale
effects. This can take into the realm of simple models such as
sand pile models, where large avalanches are produced from simple
local rules, or further down the road to agent based modelling in
general. In these areas one is looking for power laws, or sudden
crashes and changes, an apparent cooperative macroscopic behaviour
in a system with only microscopic rules - a complex system. This
for instance has been an area dubbed econophysics (see for example
\cite{Art,PBC}). In these cases the recent advances with networks
are merely providing a wider variety of playgrounds to play on. On
the other hand if these networks are more realistic, so the
network of  stock market traders is a small world network, then
perhaps the models now have a better chance of the fitting the
data.

\subsection*{Conclusions}

I have only been able to scratch the surface of networks.  There
are many more complicated types of network, many more ways of
characterising them, and far more applications than I have been
able to cover here.  What is undoubtedly true is that the last few
years have shown that there is a much richer set of possibilities
than just the old random graph or lattice, and we have been given
many new ways to characterise them.  In some ways this is just
rationalising many ideas that had been in existence for some time,
often in areas outside physics.  Physicists ought to be careful
not to assume they are the first to tread along these paths, the
roots of networks in other areas goes a long way back. Still,
physics can bring its skills and view point along with these new
ideas. There are so many applications that there is a lot more
milage in the topic, even if most readily accessible databases
have been characterised in terms of networks by now. Modelling
real complex systems is inherently difficult, its often hard to
tell if simple models capture the essence of the real system.  In
many of the problems where networks could be applied, we are
working in areas where the data is hard to collect or hard to
judge its quality. Trying to study sexually transmitted diseases
requires a knowledge of the network of sexual contacts
\cite{LEASA}.  The network as reported by academic surveys is not
going to produce data where we can estimate the errors very
easily. Physicists must also be prepared to learn new ways when
moving into other areas.

For instance my favourite real world network study is of the
social network in Marvel comic characters \cite{AMJR} in which
characters are the vertices, linked if they have appeared in the
same book. I admit, this was at first because it seemed to be an
amusing comment on how far physicists will go to jump on a
bandwagon, and perhaps it is no more ludicrous than many other
attempts to apply these network ideas. However, other academic
areas might find this work of interest even if I don't appreciate
it. After all, while we might think that many of our fictional
inventions reflect aspects of our life, even those set in some
world that has never existed, has any one ever been able to
quantify that?  Even if as a physicist I was never worried by such
a question, it might be that we can help others answer their
questions with our new network tools if we can only keep an open
mind.

\subsubsection*{After this article}

The basic ideas and computing tools are not too difficult to pick
up. There are now a good mixture of popular sources and more
technical books, reviews and papers, and a search for the word
network should turn more many more good sources than I have been
able to include. I personally found the first book of Watts
\cite{Watts99}, the review of Dorogovtsev and Mendes \cite{DM01},
the collection of papers edited by Bornholdt and Schuster
\cite{BS02} and Doug White's web site \cite{White} to be excellent
sources. There are numerous computer tools and libraries, many
free. I've used the JUNG and LEDA libraries \cite{JUNG,LEDA} and
the tools Pajek and UCINET \cite{Pajek,UCINET} and these were used
to obtain, analyse and display results for many of the models used
in the figures here.

\subsection*{Acknowledgements}

The walk algorithm for scale free networks was developed together
with Seb Klauke and Daniel Hook, and the statistical approach to
networks generation was discussed with JB Lalo\"e and Christian
Lunkes. I wish to thank them and several project students for
their discussions over the last few years. On a wider scale I have
greatly benefited by discussions with members of the ISCOM
collaboration and I thank David Lane, Sander van der Leuuw and
Geoffrey West for providing me with the opportunity to participate
as well as many useful conversations.  Without ISCOM I would not
have been able to have my horizons broadened by Doug White and
Carl Knappett and by all the other participants. I also thank the
Santa Fe Institute for support during some of these discussions.
Finally I'd like to thank Daniel Hook, Judit L\'evai, Carl
Knappett, Ray Rivers, and Dani\`ele Steer for comments on the
manuscript.

% ***************************************************************

\end{document}